\documentclass[a4paper,fleqn]{cas-sc}
\usepackage[numbers,sort&compress]{natbib}

\usepackage{amsmath,amssymb,amsfonts}
\usepackage{graphicx}
\usepackage{textcomp}
\usepackage{xcolor}
\def\BibTeX{{\rm B\kern-.05em{\sc i\kern-.025em b}\kern-.08em
		T\kern-.1667em\lower.7ex\hbox{E}\kern-.125emX}}
\usepackage[ruled,vlined,linesnumbered]{algorithm2e}
\usepackage{mathtools}
\usepackage[T1]{fontenc}
\usepackage{url}
\usepackage{dsfont}
\usepackage{color}
\usepackage{bm}
\usepackage{multirow}
\usepackage{tudacolors}
\usepackage{diagbox}
\usepackage{framed}
\usepackage{enumerate}
\usepackage{orcidlink}

\usepackage[tight]{subfigure}	

\usepackage{pgfplotstable}
\usepackage{pgfplots}
\pgfplotsset{
	compat=newest,
	every tick label/.append style={font=\tiny},
	every axis plot/.append style={line width=0.5pt, mark size=1.5pt},	
	label style={font=\color{white!15!black}, font=\scriptsize},
	legend style={legend cell align=left, align=left, draw=none, font=\scriptsize, row sep=-2pt},
}
\usetikzlibrary{plotmarks}
\usetikzlibrary{arrows.meta}
\usepgfplotslibrary{patchplots}
\usepackage{grffile}

\definecolor{MATLABcolor1}{rgb}{0.00000,0.44700,0.74100}%
\definecolor{MATLABcolor2}{rgb}{0.85000,0.32500,0.09800}%
\definecolor{MATLABcolor3}{rgb}{0.92900,0.69400,0.12500}%
\definecolor{MATLABcolor4}{rgb}{0.49400,0.18400,0.55600}%
\definecolor{MATLABcolor5}{rgb}{0.46600,0.67400,0.18800}%
\definecolor{MATLABcolor6}{rgb}{0.30100,0.74500,0.93300}%
\definecolor{MATLABcolor7}{rgb}{0.63500,0.07800,0.18400}%

\pgfplotscreateplotcyclelist{MATLABcolorlist}{
	MATLABcolor1, MATLABcolor2, MATLABcolor3, MATLABcolor4, MATLABcolor5, MATLABcolor6, MATLABcolor7
}

\pgfplotscreateplotcyclelist{MATLABcolormarklist}{
	{MATLABcolor1, mark=o, every mark/.append style = {solid}}, 
	{MATLABcolor2, mark=diamond, every mark/.append style = {solid}}, 
	{MATLABcolor3, mark=triangle, every mark/.append style = {solid,rotate=180}}, 
	{MATLABcolor4, mark=star, every mark/.append style = {solid}}, 
	{MATLABcolor5, mark=square, every mark/.append style = {solid}},
	{MATLABcolor6, mark=x, every mark/.append style = {solid}},
	{MATLABcolor7, mark=triangle, every mark/.append style = {solid}}
}

\usetikzlibrary{math}
\usetikzlibrary{calc}
\usetikzlibrary{shapes.misc}

\pgfplotsset{
	table/search path={results, images},
	table/col sep=tab,
	unbounded coords=jump,
}

\pgfplotsset{every axis/.append style={
  width=0.4\linewidth, 
}}

\tikzset{every node/.append style={font=\normalfont}}

\makeatletter
\newcommand{\IfTableHasRows}[2]{%
	\IfFileExists{\detokenize{#1}}{#2}{}%
}
\makeatother

\makeatletter
\let\orig@label\label
\renewcommand{\label}[1]{%
	\edef\@tempa{#1}%
	\ifx\@tempa\@empty
	\else
		\orig@label{#1}%
	\fi
}
\makeatother

\newcommand{\vect}[1]{\ensuremath{{\bm{#1}}}}

\newcommand\bmu{\ensuremath{\boldsymbol{\mu}}}
\newcommand\bnu{\ensuremath{\boldsymbol{\nu}}}

\newcommand{\Real}{\ensuremath{\mathbb{R}}}
\newcommand{\Compl}{\ensuremath{\mathbb{C}}}
\newcommand{\Sym}{\ensuremath{\mathbb{S}}}

\newcommand{\DiagSet}{\ensuremath{\mathbb{D}}}

\newcommand{\set}[1]{\ensuremath{\mathcal{#1}}}
\newcommand{\opr}[1]{\ensuremath{\mathcal{#1}}}

\DeclareMathOperator{\Diag}{Diag}
\DeclareMathOperator{\tr}{tr}

\DeclareMathOperator{\argmin}{argmin}
\newcommand\imagunit{\ensuremath{\mathrm{j}}}
\newcommand\euler{\ensuremath{\mathrm{e}}}

\DeclarePairedDelimiter{\norm}{\lVert}{\rVert}
\DeclarePairedDelimiter{\abs}{\lvert}{\rvert}

\newcommand{\Trans}{\ensuremath{{\mathsf{T}}}}
\newcommand{\Herm}{\ensuremath{{\mathsf{H}}}}
\newcommand{\Frob}{\ensuremath{\mathsf{F}}}

\newcommand{\st}{\text{s.t.}}

\newif\iflong
\longtrue

\begin{document}
\let\WriteBookmarks\relax
\def\floatpagepagefraction{1}
\def\textpagefraction{.001}


\shorttitle{Gridless Parameter Estimation in PCRAs}

\shortauthors{Liu et~al.}

\title [mode = title]{Gridless Parameter Estimation in Partly Calibrated Rectangular Arrays}

\tnotemark[1]
\tnotetext[1]{Part of this work was presented at IEEE ICASSP 2024~\cite{liuGridlessParameterEstimation2024}.}

\author[1]{Tianyi Liu}[orcid=0000-0001-8338-1651]
\cormark[1]
\fnmark[1]
\ead{t.liu-6@tudelft.nl}
\affiliation[1]{organization={Communication Systems Group, Technical University of Darmstadt},
            addressline={Merckstr. 25},
            city={Darmstadt},
            postcode={64283},
            country={Germany}}

\author[2]{Sai Pavan Deram}[orcid=0000-0001-5988-5534]
\ead{sai.deram@imdea.org}
\affiliation[2]{organization={IMDEA Networks Institute}, city={Madrid}, country={Spain}}

\author[3]{Khaled Ardah}[orcid=0000-0002-2396-2224]
\ead{kardah@lenovo.com}
\affiliation[3]{organization={Lenovo Research 5G Lab}, country={Germany}}

\author[4]{Martin Haardt}[orcid=0000-0001-7810-975X]
\ead{martin.haardt@tu-ilmenau.de}
\affiliation[4]{organization={Communications Research Laboratory, Ilmenau University of Technology},
            addressline={P.O. Box 10 05 65},
            city={Ilmenau},
            postcode={98684},
            country={Germany}}

\author[5]{Marc E. Pfetsch}[orcid=0000-0002-0947-7193]
\ead{pfetsch@mathematik.tu-darmstadt.de}
\affiliation[5]{organization={Research Group Optimization, Technical University of Darmstadt},
            addressline={Dolivostr. 15},
            city={Darmstadt},
            postcode={64293},
            country={Germany}}

\author[1]{Marius Pesavento}[orcid=0000-0003-3395-2588]
\ead{pesavento@nt.tu-darmstadt.de}

\fntext[1]{Present address: Signal Processing Systems Group, Delft University of Technology, Mekelweg 4, 2628 CD, Delft, The Netherlands.}

\cortext[1]{Corresponding author}


\begin{abstract}
Spatial frequency estimation from a mixture of noisy sinusoids finds applications in various fields. While subspace-based methods offer cost-effective super-resolution parameter estimation, they demand precise array calibration, posing challenges for large arrays. In contrast, sparsity-based approaches outperform subspace methods, especially in scenarios with limited snapshots or correlated sources. This study focuses on direction-of-arrival (DOA) estimation using a partly calibrated rectangular array with fully calibrated subarrays. A gridless sparse formulation is developed that leverages shift invariances in the array, yielding two competitive algorithms under the alternating direction method of multipliers (ADMM) and successive convex approximation frameworks. Numerical simulations show a superior error performance of our proposed methods, particularly in highly correlated scenarios, compared to the conventional subspace-based methods. Furthermore, we provide a generalization of the proposed method to a more challenging case where part of the sensors are unobservable due to failures.
\end{abstract}




\maketitle


\section{Introduction}
\label{sec:intro}

Direction-of-arrival (DOA) estimation is a fundamental problem in array signal processing with applications in radar, sonar, wireless communications, and acoustics~\cite{krimTwoDecadesArray1996,vantreesOptimumArrayProcessing2002,zoubirArrayStatisticalSignal2014,pesaventoThreeMoreDecades2023}. Traditional subspace-based methods such as MUSIC~\cite{schmidtMultipleEmitterLocation1986} and ESPRIT~\cite{royESPRITestimationSignalParameters1989,haardtUnitaryESPRITHow1995} offer super-resolution at low computational cost but require precise array calibration, which becomes challenging for large arrays~\cite{friedlanderSensitivityAnalysisMUSIC1990,ngArrayShapeCalibration1992}. 

Partly calibrated arrays (PCAs) address this by partitioning the array into fully calibrated subarrays with intersubarray uncertainties~\cite{liaoReviewDirectionFinding2014,pesaventoDirectionFindingPartly2002}. PCAs reduce calibration overhead while maintaining high resolution, making them attractive for mobile platforms such as UAVs and autonomous vehicles~\cite{zhangDirectionFindingPartly2023}.

Recently, several DOA estimation methods for PCAs have been developed~\cite{royESPRITestimationSignalParameters1989, parvaziDirectionofarrivalEstimationArray2011, pesaventoDirectionFindingPartly2002, seeDirectionofarrivalEstimationPartly2004}. Search-free subspace-based methods, such as ESPRIT~\cite{royESPRITestimationSignalParameters1989}, RARE~\cite{pesaventoDirectionArrivalEstimation2001,pesaventoDirectionFindingPartly2002} and the algorithm proposed in~\cite{parvaziNewDirectionofarrivalEstimation2011}, can be employed for PCAs with identical subarrays. 
Several of the above methods have also been extended to PCAs with arbitrary subarray geometries in a search-based manner~\cite{seeDirectionofarrivalEstimationPartly2004,parvaziDirectionofarrivalEstimationArray2011}. For example, a spectral search-based extension of RARE is proposed in~\cite{seeDirectionofarrivalEstimationPartly2004}. 
However, subspace-based methods typically suffer from performance degradation in difficult scenarios, e.g., in the cases with highly correlated source signals and/or a limited number of measurements. To address those challenges, sparsity-based methods have been explored for PCAs~\cite{steffensDirectionFindingArray2014,steffensCompactFormulationEll2018}. In \cite{steffensBlockRanksparseRecovery2018}, the authors employ a grid-based sparse model and estimate DOAs by solving a block- and rank-sparse optimization problem. Whereas grid-based methods may suffer from basis mismatch due to the spectrum discretization, a gridless sparse method, termed shift-invariant SPARROW (SI-SPARROW), is introduced in~\cite{steffensGridlessCompressedSensing2017} for PCAs, avoiding the sampling over the field-of-view (FOV). Despite the promising error performance demonstrated by the above sparse methods even in challenging scenarios, they both assume a uniform linear subarray structure, limiting DOA estimation to the azimuth direction.
The extreme case with only a single snapshot in PCAs is studied in~\cite{zhangDirectionFindingPartly2023}, and a novel method that achieves an enhanced angular resolution is proposed by exploiting the approximate orthogonality of the source signals after a transformation.

The exact synchronization of distributed arrays involves the calibration of both amplitude and phase responses within each subarray and between subarrays. Here, we focus on the case of only intersubarray phase uncertainties and the amplitude responses are assumed to be perfectly calibrated.
More general studies that address both amplitude and phase errors can be found in~\cite{zhangDirectionFindingPartly2023} and the references therein.

This paper extends the gridless SI-SPARROW formulation~\cite{steffensGridlessCompressedSensing2017} to partly calibrated rectangular arrays (PCRAs), enabling 2D DOA estimation in both azimuth and elevation.
Although, for notational simplicity, the SI-SPARROW formulation is presented for a PCRA of identical subarrays, it can readily be applied to more general array geometries by employing other shift invariances and/or including virtual unobservable sensors as described in Section~\ref{sec:unobservable}.
Given the solution of the SI-SPARROW formulation, a multidimensional ESPRIT-like method based on the simultaneous diagonalization~\cite{fuSimultaneousDiagonalizationSimilarity2006} is then provided to finally recover and automatically pair the azimuth and elevation DOAs.
Numerical simulations demonstrate the robustness of our proposed method to the source correlation, compared to the conventional approach that applies the ESPRIT-like methods directly to the sample covariance matrix. 

Moreover, in contrast to the semidefinite programming (SDP) reformulation approach in~\cite{steffensGridlessCompressedSensing2017}, we develop two competitive algorithms under the alternating direction method of multipliers (ADMM) and the successive convex approximation (SCA) frameworks for the established SI-SPARROW problem.
The proposed algorithms can be parallelized and exhibit a significantly reduced computation time with the increase of the sample size and/or array size, compared to the corresponding SDP implementation.
In practical scenarios, the customized parallel algorithms become indispensable as the SDP implementation is computationally unattainable.

Naturally, the SI-SPARROW formulation can also be employed in the fully calibrated case as a low-cost decorrelation step to improve the robustness of the subspace-based methods, e.g., MUSIC, to source correlation.

To summarize, the main contributions of this paper are:
\begin{itemize}
  \item We introduce a gridless sparse formulation for DOA estimation in a PCRA composed of fully calibrated subarrays by leveraging shift invariances.
    From the simulation results, the proposed formulation shows a superior error performance compared to the conventional subspace methods, particularly in highly correlated scenarios.
  \item We develop two complementary parallel algorithms under the ADMM and SCA frameworks for the established SI-SPARROW problem that do not require the use of inefficient general-purpose solvers.
    Each of them has merit in terms of computational cost in different scenarios.
  \item We also provide a generalization of the proposed method to a more challenging case with unobservable sensors.
\end{itemize}

The paper is organized as follows.
Section~\ref{sec:model} introduces the signal model and the shift-invariance properties.
Section~\ref{sec:problem} presents the SI-SPARROW formulation.
In Sections~\ref{sec:ADMM} and~\ref{sec:SCA}, we devise two competitive algorithms for the SI-SPARROW problem under the ADMM and SCA frameworks.
Section~\ref{sec:unobservable} discusses the generalization to unobservable sensors.
Simulation results and conclusions follow in Sections~\ref{sec:results} and~\ref{sec:conclusion}.

\textit{Notation:}
We use $ x $, $ \vect{x} $, and $ \vect{X} $ to denote a scalar, column vector, and matrix, respectively.
The $k$th entry of a vector $\vect{x}$ is denoted by $x_k$ and the $(k,l)$th entry of a matrix $\vect{X}$ is $x_{k,l}$.
For any $x \in \Compl$, $x^*$ denotes its complex conjugate.
For matrix $\vect{X}$, $ \vect{X}^\Trans $, $ \vect{X}^\Herm $, $ \vect{X}^{-1} $, $ \tr (\vect{X}) $, and $ \norm{\vect{X}}_\Frob $ denote the transpose, Hermitian transpose, inverse, trace, and Frobenius norm of $\vect{X}$, respectively.
The sets of $ M \times M $ diagonal, nonnegative diagonal, Hermitian, and positive semidefinite (PSD) Hermitian matrices are denoted by $\DiagSet^M$, $\DiagSet_+^M$, $ \Sym^M $ and $ \Sym_+^M $, respectively. The $M \times M$ identity matrix is denoted by $\vect{I}_M$, and $\vect{0}$ and $\vect{1}$ represent a zero matrix and all-ones matrix, respectively.
The Euclidean projection onto a set $ \set{X} $ is denoted by $ \opr{P}_{\set{X}}(\cdot) $.
We define the indicator function of a closed set $\set{X}$ as
$ \mathbb{I}_{\set{X}} (\vect{x}) = 0 $ for $ \vect{x} \in \set{X} $ and $ \mathbb{I}_{\set{X}} (\vect{x}) = + \infty $ otherwise.

\section{Signal Model}
\label{sec:model}

Consider a PCRA, as shown in Fig.~\ref{fig:PCRA}, composed of $P$ identical rectangular subarrays of $L_{x} \times L_{y}$ sensors where the sensors within each subarray are well calibrated, but not the intersubarray elements. 
Let $M= P L_x L_y$ be the number of sensors in the PCRA. 
We assume that $N_\text{S}$ narrowband far-field source signals impinge from distinct unknown DOAs with azimuth and elevation angles denoted by $ \phi_i \in [-180^\circ, 180^\circ) $ and $ \theta_i \in [0^\circ, 90^\circ] $, respectively, for $i = 1, \ldots, N_\text{S}$. Each direction $ (\phi_i,\theta_i) $ can be equivalently represented by a pair of spatial frequencies in the two dimensions defined as
\begin{equation}
	\mu_{i}^x = \pi \cos(\phi_{i}) \sin(\theta_{i}) \quad \text{and} \quad \mu_{i}^y = \pi \sin(\phi_{i}) \sin(\theta_{i}),
\end{equation}
respectively. The spatial frequencies of the $ N_\text{S} $ sources are summarized in $ \bmu \!=\! [(\bmu^x)^\Trans, (\bmu^y)^\Trans]^\Trans $ with $ \bmu^x \!=\! [\mu_1^x,\ldots,\mu_{N_\text{S}}^x]^\Trans $ and $ \bmu^y \!=\! [\mu_1^y,\ldots,\mu_{N_\text{S}}^y]^\Trans $.
Let $ \vect{Y} \! \in \! \Compl^{M \times N} $ be the measurement matrix 
where the $ (m,n) $th entry $ y_{mn} $ is the output at sensor $ m $ in time-slot $ n $. The measurement matrix is modeled as
\begin{equation}\label{eq:model}
  \vect{Y} = \vect{A} (\bmu, \vect{\alpha})   \vect{\Psi} + \vect{N}, 
\end{equation}
where $\vect{\Psi} \in \Compl^{N_\text{S} \times N}$ contains the source waveforms with $ \psi_{i,n} $ being the waveform from source $ i $ at time instant $ n $. 
The matrix $\vect{N} \in \Compl^{M \times N}$ comprises noise entries independently and identically distributed (i.i.d.) according to the Gaussian distribution $\mathcal{CN}(0,\sigma_n^2)$ and $\vect{A} (\bmu, \vect{\alpha}) \in \mathbb{C}^{M\times N_\text{S}}$ is the matrix collecting the $N_\text{S}$ steering vectors of the 2D array as 
\begin{equation}\label{eq:steerMat}
  \vect{A} (\bmu,\vect{\alpha}) = \begin{bmatrix}
    \vect{a}(\mu_{1}^x ,\mu_{1}^y, \vect{\alpha}) & \ldots & \vect{a}(\mu_{N_\text{S}}^x ,\mu_{N_\text{S}}^y, \vect{\alpha})
	\end{bmatrix},
\end{equation}
where $\vect{a}(\mu_{i}^x,\mu_{i}^y,\vect{\alpha}) \in \Compl^{M}$ is the array-dependent steering vector in the direction of source $ i $.
For the considered PCRA, the steering vector in the direction $ (\mu^x,\mu^y) $ is expressed as 
\begin{equation}\label{eq:steerVec}
  \vect{a}(\mu^x ,\mu^y,\vect{\alpha}) = \vect{h}(\mu^x, \mu^y, \vect{\alpha}) \otimes \vect{v} (\mu^x,\mu^y).
\end{equation}
The vector $ \vect{v}(\mu^x,\mu^y) \in \Compl^{L_x L_y} $ describes the steering vector of each subarray, i.e., the calibrated part of the array manifold, whereas $ \vect{h} (\mu^x, \mu^y, \vect{\alpha}) \in \Compl^P $ characterizes the unknown perturbations between subarray responses, i.e., the uncalibrated part, which depends not only on the frequencies $(\mu^x, \mu^y)$ but also the unknown array parameters summarized in the vector $ \vect{\alpha} $.
For the considered rectangular subarray, which is a Cartesian product of two linear arrays, $ \vect{v}(\mu^x,\mu^y) $ can be written as
\begin{equation} \label{eq:subarraySteerVec}
  \begin{aligned}
    \vect{v}(\mu^x,\mu^y) &= \vect{v}^x (\mu^x) \otimes \vect{v}^y (\mu^y) \\
		\text{with} \quad \vect{v}^x (\mu^x) &= [1, \euler^{\imagunit \mu^x \delta^x_2}, \ldots, \euler^{\imagunit \mu^x \delta^x_{L_x}}]^\Trans, \\
		\vect{v}^y (\mu^y) &= [1, \euler^{\imagunit \mu^y \delta^y_2}, \ldots, \euler^{\imagunit \mu^y \delta^y_{L_y}}]^\Trans,
	\end{aligned}
\end{equation}
where $\delta^{x}_{l}$ (resp., $\delta_l^y$) is the known relative position of the $l$th sensor in the $x$-axis (resp., $y$-axis) within each subarray.
In this work, we focus on the case where the intersubarray displacements and phase-offsets are uncalibrated.
Thus, the uncalibrated part of the steering vector in~\eqref{eq:steerVec} is expressed as
\begin{equation} \label{eq:uncalibratedSteerVec}
  \vect{h} (\mu^x, \mu^y, \vect{\alpha}) = \Big[1, \ \euler^{\imagunit \left( \mu^x \Delta^x_2 + \mu^y \Delta^y_2 + \tau_2 \right)}, \ \ldots, \ \euler^{\imagunit \left( \mu^x \Delta^x_P + \mu^y \Delta^y_P + \tau_P \right)}\Big]^\Trans,
\end{equation}
where $ (\Delta^x_p,\Delta^y_p) $ and $\tau_p$ are the unknown coordinates and phase-offset, respectively, of the $p$th subarray relative to the first subarray.
Moreover, both the frequencies $ \vect{\mu} $ and the unknown array parameters $ \vect{\alpha} = [\Delta^x_2,\ldots,\Delta^x_P,\Delta^y_2,\ldots,\Delta^y_P,\tau_2,\ldots,\tau_P]^\Trans $ are assumed to be constant during the measurement time.


\begin{figure}
	\centering
	\subfigure[\label{subfig:selMat_intersubarray}]{\pgfdeclarelayer{background}
\pgfdeclarelayer{foreground}
\pgfdeclarelayer{main}
\pgfsetlayers{background,main,foreground}
\begin{tikzpicture}[scale=0.5]
  \input{images/arrayParams} 
	
  \begin{pgfonlayer}{foreground}
    \input{images/sensors}
  \end{pgfonlayer}

  \begin{pgfonlayer}{background}
    \tikzmath{ 
      int \jx,\jy,\iSubarray;
      for \iSubarray in {1,...,\nSubarray}{
        \subarrayX = \subarrayXcoord\iSubarray; 
        \subarrayY = \subarrayYcoord\iSubarray; 
        { 
          \draw[thick,draw=TUDa-1a,fill=TUDa-0a, rounded corners] (\x0+\subarrayX-0.5,\y0+\subarrayY-0.5) rectangle (\x0+\subarrayX+\nSensorX*\spaceX-\spaceX+0.5,\y0+\subarrayY+\nSensorY*\spaceY-\spaceY+0.5);
          \node[above,anchor=south east] at (\x0+\subarrayX+\nSensorX*\spaceX-\spaceX+0.5,\y0+\subarrayY+\nSensorY*\spaceY-\spaceY+0.5) {\footnotesize subarray \iSubarray };
        };
      };
      \xlast = \x0+\subarrayX+\nSensorX*\spaceX-\spaceX;	
      \ylast = \y0+\subarrayY+\nSensorY*\spaceY-\spaceY;	
      for \jx in {1,...,\nSensorX}{
        { 
          \xtick{(\x0+\jx*\spaceX-\spaceX,0)}{}{};
        };
        if \jx > 1 then {
          { 
            \draw[thick,stealth-stealth,color=TUDa-1b] (\x0,-0.5) -- (\x0+\jx*\spaceX-\spaceX,-0.5) node[TUDa-1c,below] {$ \delta_{\jx}^x $};
          };
        };
      };
      for \jy in {1,...,\nSensorY}{
        { 
          \ytick{(0,\y0+\jy*\spaceY-\spaceY)}{\iy}{\jy};
        };
        if \jy > 1 then {
          { 
            \draw[thick,stealth-stealth,color=TUDa-1b] (-0.5,\y0) -- (-0.5,\y0+\jy*\spaceY-\spaceY) node[TUDa-1c,left] {$ \delta_{\jy}^y $};
          };
        };
      };
      for \iSubarray in {1,...,\nSubarray}{
        \subarrayX = \subarrayXcoord\iSubarray; 
        \subarrayY = \subarrayYcoord\iSubarray; 
        if \iSubarray > 1 then {
          { 
            \draw[thick,color=TUDa-8b] (\x0+\subarrayX,\y0+\subarrayY) -- (\x0+\subarrayX-0.5,\y0+\subarrayY-0.5) node[below,anchor=north,color=TUDa-8b] {$ (\Delta_\iSubarray^x, \Delta_\iSubarray^y) $};
          };
        } else {
          { 
            \draw[thick,color=TUDa-8b] (\x0+\subarrayX,\y0+\subarrayY) -- (\x0+\subarrayX-0.5,\y0+\subarrayY-0.5) node[below,anchor=north,color=TUDa-8b] {$ (0,0) $};
          };
        };
      };
    }
    
    \draw[axis] (0,0) -- (\xlast+1,0) node[below] {$ x $};
    \draw[axis] (0,0) -- (0,\ylast+2) node[anchor = north east] {$ y $};
  \end{pgfonlayer}

  \begin{pgfonlayer}{main}
    \tikzmath{ 
      int \jx,\jy,\iSubarray;
      for \iSubarray in {1,...,\nSubarray}{
        \subarrayX = \subarrayXcoord\iSubarray; 
        \subarrayY = \subarrayYcoord\iSubarray; 
        { 
          \draw[thick,draw=TUDa-3a, rounded corners] (\x0+\subarrayX-0.5,\y0+\subarrayY-0.5) rectangle (\x0+\subarrayX+\nSensorX*\spaceX-\spaceX+0.5,\y0+\subarrayY+\nSensorY*\spaceY-\spaceY+0.5);
          \node[right,anchor=south west,TUDa-3c] at (\x0+\subarrayX+\nSensorX*\spaceX-\spaceX+0.5,\y0+\subarrayY+\nSensorY*\spaceY-\spaceY) {$ \vect{J}_{\iSubarray} $};
        };
      };
    }
  \end{pgfonlayer}
	
\end{tikzpicture}}
  \hfill
	\subfigure[\label{subfig:selMat_intrasubarray}]{
\pgfdeclarelayer{background}
\pgfdeclarelayer{foreground}
\pgfsetlayers{background,foreground}
\begin{tikzpicture}[scale=0.5]
	\input{images/arrayParams}
	
  \begin{pgfonlayer}{foreground}
    \input{images/sensors}
  \end{pgfonlayer}

  \begin{pgfonlayer}{background}
    \tikzmath{ 
      int \ix,\jx,\iy,\jy,\iSubarray;
      integer \color1,\color2;
      \color1 = 8; \color2 = 10; 
      coordinate \textcoord1,\textcoord2;
      \textcoord1 = (\x0+\subarrayXcoord2,\y0);
      \textcoord2 = (\x0+\subarrayXcoord2+\spaceX,\ymax+3*\spaceY);
      \groupAnchorY1 = -0.4*\spaceY;
      \groupAnchorY2 = 0.4*\spaceY;
      for \jy in {1,...,\nSensorY}{
        int \jc;
        \jc = \color\jy;
        {
          \node[color=TUDa-\jc c,below] (text) at (\textcoord\jy) {$ \vect{K}_{\jy}^y $};
        };
        for \iSubarray in {1,...,\nSubarray}{
          \subarrayX = \subarrayXcoord\iSubarray; 
          \subarrayY = \subarrayYcoord\iSubarray; 
          \sensorY = \y0+\subarrayY+\jy*\spaceY-\spaceY; 
          {
            \draw[thick,draw=TUDa-\jc a,rounded corners,dashed] (\x0+\subarrayX-0.4*\spaceX,\sensorY-0.4*\spaceY) rectangle (\x0+\subarrayX+\nSensorX*\spaceX-\spaceX+0.4*\spaceX,\sensorY+0.4*\spaceY); 
            \draw[thick,-stealth,draw=TUDa-\jc b] (\x0+\subarrayX+\spaceX,\sensorY+\groupAnchorY\jy) -- (text);
          };
        };
      };
    }
  \end{pgfonlayer}
	
\end{tikzpicture}}
	\caption{Example of a PCRA composed of $ 3 $ subarrays with $ 3 \times 2 $ sensors per subarray. The shift-invariant groups that involve the intersubarray displacements are indicated in~\subref{subfig:selMat_intersubarray}, whereas  the shift-invariant groups that involve the intrasubarray displacements in dimension $y$ are indicated in~\subref{subfig:selMat_intrasubarray}.}
	\label{fig:PCRA}
\end{figure}

In the following, we introduce the shift-invariance properties of the array. Toward this goal, we define the selection matrices
\begin{subequations} \label{eq:selectMat}
	\begin{align}
		\vect{J}_p &= \vect{e}_{P,p} \otimes \vect{I}_{L_x} \otimes \vect{I}_{L_y}, & p = 1,\ldots,P,\\ 
		\vect{K}_l^x &= \vect{I}_{P} \otimes \vect{e}_{L_x,l} \otimes \vect{I}_{L_y}, & l = 1,\ldots,L_x, \\ 
		\vect{K}_l^y &= \vect{I}_{P} \otimes \vect{I}_{L_x} \otimes \vect{e}_{L_y,l}, & l = 1,\ldots,L_y,  
	\end{align}
\end{subequations}
to assign sensors to various shift-invariant groups, where $\vect{e}_{P,p} = [0,\ldots,0,1,0,\ldots,0]^\Trans$ is 
the $p$th column of $ \vect{I}_P $. 
As depicted in Fig.~\ref{fig:PCRA}, by the operation $ \vect{J}_p^\Trans \vect{a} $, the responses of the $ p $th subarray are selected. Similarly, $ \vect{K}_l^x $ and $ \vect{K}_l^y $ are used to select all sensors at the $ l $th position in the $ x $-axis and $ y $-axis, respectively, within the subarrays.
Then the shift-invariance properties of the steering matrix are given as
\begin{subequations}\label{eq:shift-invariance-prop}
  \begin{align} 
    \vect{J}_{p}^\Trans   \vect{A} (\bmu,\vect{\alpha}) &\!=\! \vect{J}_{1}^\Trans   \vect{A} (\bmu,\vect{\alpha})  \vect{\Phi}(\Delta_{p}^{x}\bmu^x + \Delta_p^y \vect{\mu}^y + \tau_p \vect{1}), & \quad p \!=\! 2,\dots,P, \label{eq:shift-invariance_intersubarray} \\
    {(\vect{K}_{l}^{x})}^\Trans \vect{A}(\bmu,\vect{\alpha}) &\!=\! {(\vect{K}_{1}^{x})}^\Trans   \vect{A} (\bmu,\vect{\alpha})  \vect{\Phi}({\delta_{l}^{x}}\bmu^x), & \quad l \!=\! 2,\dots,L_x, \label{eq:shift-invariance_sensor_x} \\
    {(\vect{K}_{l}^{y})}^\Trans \vect{A}(\bmu,\vect{\alpha}) &\!=\! {(\vect{K}_{1}^{y})}^\Trans   \vect{A} (\bmu,\vect{\alpha})  \vect{\Phi}({\delta_{l}^{y}}\bmu^y), & \quad l \!=\! 2,\dots,L_y, \label{eq:shift-invariance_sensor_y}
  \end{align}
\end{subequations}
where $\vect{\Phi}(\vect{x}) \!=\! \Diag (\euler^{\imagunit x_{1}},\ldots,\euler^{\imagunit x_{N_\text{S}}}) \! \in \! \Compl^{N_\text{S} \times N_\text{S}} $ for a vector $\vect{x} \in \Real^{N_\text{S}}$.
Also, it is straightforward that each single sensor forms a shift-invariant group regardless of the array configuration. 
The shift invariance between the first sensor of the first subarray and the $(k,l)$th sensor of the $p$th subarray is written as
\begin{equation}\label{eq:shift-single-sensor}
  \vect{e}_{M,m}^\Trans \vect{A} (\bmu,\vect{\alpha}) = \vect{e}_{M,1}^\Trans \vect{A} (\bmu,\vect{\alpha}) \vect{\Phi} \big({(\Delta_p^x + \delta_k^x)}\bmu^x + {(\Delta_p^y + \delta_l^y)}\bmu^y + \tau_p \vect{1} \big)
\end{equation}
for $ p = 1,\ldots,P$, $k = 1,\ldots,L_x, \ l = 1,\ldots,L_y $ and $ m = (p-1)L_x L_y + (k-1)L_y + l $.
Additional shift invariances can be exploited for other array topologies but, for simplicity, we focus on the example of the PCRA in Fig.~\ref{fig:PCRA} in this paper.





\section{Sparse signal formulation}
\label{sec:problem}


This section begins with an introduction to an equivalent sparse representation of the model in~\eqref{eq:model} obtained by sampling the FOV and the state-of-the-art SPARROW formulation~\cite{steffensCompactFormulationEll2018} constructed on the grid-based sparse model.
To address intersubarray uncertainties, a gridless relaxation of the SPARROW formulation is proposed by using the shift invariances.

\subsection{Sparse Formulation for Fully Calibrated Arrays}
\label{subsec:SPARROW}
We first review the classic grid-based sparse formulation for DOA estimation, which requires the full calibration.

We define a sparse representation for the model in~\eqref{eq:model} as
\begin{equation} \label{eq:sparseModel}
  \vect{Y} = \vect{A} (\bnu, \vect{\alpha} ) \vect{X} + \vect{N},
\end{equation}
where $\vect{A} (\bnu,\vect{\alpha} ) \in \Compl^{M \times K}$ is an overcomplete dictionary constructed according to~\eqref{eq:steerMat} by sampling the field-of-view in $K \gg N_\text{S}$ directions with spatial frequencies $ \bnu = [{(\bnu^x)}^\Trans, {(\bnu^y)}^\Trans]^\Trans $, where $ \bnu^x = [\nu_1^x,\ldots,\nu_K^x]^\Trans $ and $ \bnu^y = [\nu_1^y,\ldots,\nu_K^y]^\Trans $, and $ \vect{X} \in \Compl^{K \times N} $ is a sparse representation of the waveform matrix $ \vect{\Psi} $. 
Provided that the true frequencies $ \bmu $ are contained in the frequency grid, i.e.,
\begin{equation} \label{eq:on-grid}
	\{ (\mu_i^x,\mu_i^y)\}_{i=1}^{N_\text{S}} \subset \{(\nu_k^x, \nu_k^y)\}_{k=1}^K,
\end{equation}
$\vect{X} = [\vect{x}_1,\ldots,\vect{x}_K]^\Trans$ has only $N_\text{S}$ nonzero rows $ \vect{x}_k^\Trans $ corresponding to the waveforms of the sources.
For simplicity, in the rest of the paper, the dictionary is denoted as $ \vect{A} = \vect{A} (\bnu,\vect{\alpha}) $.
Based on~\eqref{eq:sparseModel}, the DOA estimation problem is formulated as the well-known convex mixed-norm minimization
\begin{equation}\label{prob:mixed-norm}
	\min_{\vect{X} \in \Compl^{K \times N}} \quad \tfrac{1}{2} \norm{\vect{Y} - \vect{A} \vect{X}}_\Frob^{2} + \lambda \sqrt{N}\norm{\vect{X}}_{2,1},
\end{equation}
where $\lambda > 0$ is the regularization parameter and
		$\norm{\vect{X}}_{2,1} = \sum_{k = 1}^{K} \norm{\vect{x}_k}_{2}$
is the $\ell_{2,1}$-norm.
As we are only interested in the DOAs, problem~\eqref{prob:mixed-norm} can be equivalently reformulated as the SPARse ROW-norm reconstruction (SPARROW) problem~\cite{steffensCompactFormulationEll2018}
\begin{equation}\label{prob:gridSPARROW}
	\min_{{\vect{S}}\in \DiagSet_{+}^K} \quad \tr { \big((\vect{A}\vect{S}\vect{A}^\Herm + \lambda \vect{I}_{M})^{-1} \widehat{\vect{R}} \big)}+\tr(\vect{S})
\end{equation}
with the sample covariance matrix $\widehat{\vect{R}} \!=\! \vect{YY}^\Herm / N$.
The solutions $ \widehat{\vect{X}} $ and $ \widehat{\vect{S}} $ of problems~\eqref{prob:mixed-norm} and~\eqref{prob:gridSPARROW}, respectively, are related by
$ \widehat{\vect{S}} = {1}/{\sqrt{N}}\Diag (\norm{\widehat{\vect{x}}_{1}}_{2},\ldots,\norm{\widehat{\vect{x}}_{K}}_{2}) $.	
Compared to~\eqref{prob:mixed-norm}, problem~\eqref{prob:gridSPARROW} has a reduced variable size, with DOAs indicated by the nonzero diagonal entries of $\vect{S}$. It can be reformulated as a SDP solvable by generic solvers, e.g., MOSEK~\cite{mosekapsMOSEKOptimizationToolbox2023}, or more efficiently by a coordinate descent algorithm~\cite{steffensCompactFormulationEll2018} for large arrays.

However, the grid-based formulation~\eqref{prob:gridSPARROW} has two drawbacks.
First, the on-grid assumption~\eqref{eq:on-grid} is often impractical due to the limited grid size, leading to spectral leakage and basis mismatch~\cite{chiSensitivityBasisMismatch2011,hermanGeneralDeviantsAnalysis2010} in the recovered signal.
Second, constructing the dictionary $ \vect{A} $ requires complete array calibration. 

A natural gridless extension of~\eqref{prob:gridSPARROW} is jointly learning a dictionary $ \vect{A} $ in the array manifold, i.e.,
\begin{equation}\label{prob:jointSPARROW}
	\min_{{\vect{S}}\in \DiagSet_{+}^K,\, \vect{A} \in \set{A}^K} \quad \tr { \big((\vect{A}\vect{S}\vect{A}^\Herm + \lambda \vect{I}_{M})^{\rm -1} \widehat{\vect{R}} \big)}+\tr(\vect{S}),
\end{equation}
where $ \set{A}^K \!=\! \{ \vect{A}(\bnu,\vect{\alpha}) \! \mid \! \bnu \! \in \! [-\pi, \pi]^{2K}, (\nu_i^x \!, \nu_i^y) \! \neq \! (\nu_j^x,\nu_j^y) \ \forall i,j \!=\! 1,\ldots,K, \, i \! \neq \! j \} $ is the array manifold with $ K $ distinct DOAs.
As the steering vectors 
contain unit-modulus entries,
we have
	$	\tr(\vect{ASA}^\Herm) 
  = M \tr(\vect{S})$.
Thus, with a slack variable $ \vect{Q} $, problem~\eqref{prob:jointSPARROW} can be equivalently written as
\begin{subequations} \label{prob:jointSPARROWreform}
	\begin{align}
		\min_{{\vect{S}}\in \DiagSet_{+}^K, \vect{A} \in \set{A}^K, \vect{Q} \in \Sym_+^M} & \ M \tr \big(({\vect{Q}} + \lambda \vect{I}_{M})^{-1}\widehat{\vect{R}} \big)+\tr (\vect{Q}) \label{eq:jointSPARROWreform_obj}\\
		\st & \ \vect{Q} = \vect{A} \vect{S} \vect{A}^\Herm. \label{eq:dictConstraint}
	\end{align}
\end{subequations}
The objective function in~\eqref{prob:jointSPARROWreform} depends only on $ \vect{Q} $ whose structure is specified by the dictionary-based constraint~\eqref{eq:dictConstraint}. 
The trace-term $ \tr (\vect{Q}) $ in~\eqref{eq:jointSPARROWreform_obj} promotes rank-sparsity of $ \vect{Q} $ as it is equivalent to the nuclear norm for $ \vect{Q} \succeq 0 $~\cite{candesExactMatrixCompletion2009}.

For a uniform linear array (ULA), the dictionary $\vect{A}$ has a Vandermonde structure, which results in a Toeplitz structure on $ \vect{Q} $.
The Carath\'eodory theorem~\cite{caratheodoryUberVariabilitatsbereichFourierschen1911,caratheodoryUberZusammenhangExtremen1911,toeplitzZurTheorieQuadratischen1911} ensures the existence of a unique Vandermonde decomposition in the form of~\eqref{eq:dictConstraint} for any rank-deficient PSD Toeplitz matrix $\vect{Q}$.
Hence, the dictionary-specific structure of $\vect{Q}$ in~\eqref{eq:dictConstraint} can be equivalently replaced by a Toeplitz structure and the resulting formulation is termed gridless SPARROW (GL-SPARROW)~\cite{steffensCompactFormulationEll2018}. 
From the minimizer $ \widehat{\vect{Q}} $, the DOAs preserved in its Vandermonde decomposition can be recovered by any fully-calibrated methods, e.g., root-MUSIC, due to the uniqueness of the decomposition.


Multilevel Toeplitz structures have been exploited for arrays of a multilevel uniform structure, e.g., 2D uniform rectangular arrays (URAs)~\cite{yangVandermondeDecompositionMultilevel2016,zhangEfficientTwodimensionalLine2019}.
An analog of the Vandermonde decomposition for multilevel Toeplitz matrices has been discovered, and this decomposition is unique when the rank of the matrix is sufficiently low~\cite{yangVandermondeDecompositionMultilevel2016}.

For a general array geometry, problem~\eqref{prob:jointSPARROWreform}, however, cannot be easily solved and the construction of the array manifold still requires complete array calibration. A simple gridless relaxation may be obtained by completely discarding the dictionary-specific structural constraint~\eqref{eq:dictConstraint},
which equivalently relaxes the array manifold $ \set{A}^K $ to be the complete space $ \Compl^{M \times K} $ and, consequently, leads to a meaningless solution.

\subsection{Gridless Shift-Invariant Formulation for PCRAs}
\label{subsec:SI-SPARROW}

A generalization of GL-SPARROW is proposed in~\cite{steffensGridlessCompressedSensing2017} for partly calibrated linear arrays by replacing the Toeplitz structure by less rigid structural constraints derived from shift invariances.
In the following, we extend this approach to the PCRA in Fig.~\ref{fig:PCRA}.
Applying the shift-invariance properties in~\eqref{eq:shift-invariance-prop} to the dictionary $\vect{A}(\bnu,\vect{\alpha})$, we obtain the following identity:
\begin{align*}
    & {(\vect{K}_{l}^x)}^\Trans \vect{Q} \vect{K}_{l}^x = {(\vect{K}_{l}^x)}^\Trans   \vect{A} \vect{S} \vect{A}^\Herm   \vect{K}_{l}^x
    = {(\vect{K}_{1}^x)}^\Trans   \vect{A}   \vect{\Phi}({\delta_{l}^{x}}\bnu^x) \vect{S} \vect{\Phi}^\Herm({\delta_{l}^{x}}\bnu^x) \vect{A}^\Herm \vect{K}_{1}^x 
     = {(\vect{K}_{1}^x)}^\Trans \vect{Q}  \vect{K}_{1}^x
\end{align*}
for $ l = 2,\ldots,L_x $.
The same identity can be extended to the other shift-invariance groups stated in~\eqref{eq:shift-invariance-prop} and~\eqref{eq:shift-single-sensor}, which leads to the following set of structural constraints:
\begin{subequations}\label{eq:shift-invariance-constraint}
	\begin{align}
    \vect{J}_{p}^\Trans   \vect{Q}   \vect{J}_{p} &= \vect{J}_{1}^\Trans \vect{Q}  \vect{J}_{1}, & p = 2,\ldots,P, \label{eq:shift-invariance-constraint_intersubarray}\\
		{(\vect{K}_{l}^{x})}^\Trans   \vect{Q}   \vect{K}_{l}^{x} &= {(\vect{K}_{1}^{x})}^\Trans \vect{Q}  \vect{K}_{1}^{x}, & l = 2,\ldots,L_x,\\
		{(\vect{K}_{l}^{y})}^\Trans   \vect{Q}   \vect{K}_{l}^{y} &= {(\vect{K}_{1}^{y})}^\Trans \vect{Q}  \vect{K}_{1}^{y}, & l = 2,\ldots,L_y, \\
		q_{ii} &= q_{11}, & i= 2,\ldots,M.
	\end{align}
\end{subequations}
The constraints~\eqref{eq:shift-invariance-constraint} define a subspace of $M \! \times \! M$ Hermitian matrices, which we denote by
\begin{equation} \label{eq:shift-invariant-subspace}
	\set{T}^M = \{\vect{Q} \in \Sym^M \mid \vect{Q} \text{ satisfies the constraints in~\eqref{eq:shift-invariance-constraint}}\}.
\end{equation}
Each equality in~\eqref{eq:shift-invariance-constraint} defines a pair of principal submatrices of $\vect{Q}$ to be identical. This implies that the shift-invariant structure defined by~\eqref{eq:shift-invariance-constraint} is a generalization of the multilevel Toeplitz structure.
Replacing the dictionary-based constraint~\eqref{eq:dictConstraint} by the structural constraints~\eqref{eq:shift-invariance-constraint} leads to the following gridless shift-invariant SPARROW (SI-SPARROW) problem:
\begin{equation} \label{prob:SI-SPARROW}
	\min_{\vect{Q} \in \Sym_+^M \cap \set{T}^M} \quad f(\vect{Q}) = M \tr \big(({\vect{Q}} + \lambda \vect{I}_{M})^{-1}\widehat{\vect{R}} \big)+\tr (\vect{Q}).
\end{equation}
The shift-invariant structure does not necessarily ensure a unique decomposition in the form of~\eqref{eq:dictConstraint}. Therefore, the SI-SPARROW problem~\eqref{prob:SI-SPARROW} is generally a relaxation of~\eqref{prob:jointSPARROWreform}.

As a widely-used shift-invariance-based method, the multidimensional ESPRIT (MD-ESPRIT)~\cite{zhangChannelEstimationTraining2017,zoltowskiClosedform2DAngle1996,haardtSimultaneousSchurDecomposition1998} can be used to recover the spatial frequencies with automatic pairing from the sample covariance matrix $ \widehat{\vect{R}} $, based only on the shift invariances in~\eqref{eq:shift-invariance_sensor_x} and~\eqref{eq:shift-invariance_sensor_y} that involve the known array parameters, i.e., intrasubarray displacements $ \delta_l^x $ and $ \delta_l^y $.
According to~\eqref{eq:dictConstraint}, $ {\vect{Q}} $ is originally designed to span the same subspace as the steering matrix $\vect{A}$. 
Hence, MD-ESPRIT can also be used to further recover the spatial frequencies from the solution $ \widehat{\vect{Q}} $ of~\eqref{prob:SI-SPARROW} in a search-free manner.
Compared to $ \widehat{\vect{R}} $, the sources are decorrelated in $ \widehat{\vect{Q}} $, leading to an improved estimation performance particularly for highly correlated sources.

The number of sources $ N_{\text{S}} $ is assumed known, as in other subspace methods, and is used to determine the signal subspace of $ \widehat{\vect{Q}} $ in the subsequent MD-ESPRIT step.
Thus, the eigenvalues in the noise subspace of $ \widehat{\vect{Q}} $ only need to be small but not necessarily zero,
allowing flexibility in tuning the penalty weight $ \lambda $ for balancing the bias-variance trade-off~\cite[Sec. 6.2]{jamesIntroductionStatisticalLearning2021}. 
That is, the utilization of regularization typically leads to decreased estimation variance but increased bias.

While~\eqref{prob:SI-SPARROW} is formulated for PCRAs with identical subarrays, the approach is not limited to a specific array geometry or shift-invariant subspace $\set{T}^M$. The proposed algorithms can be readily applied to a similar problem with a subspace $\set{T}^M$ defined by other shift invariances, and the formulation can be extended to higher-dimensional arrays by including additional invariances.


Similar to the grid-based and the gridless SPARROW that exploits a Toeplitz structure, problem~\eqref{prob:SI-SPARROW} can be equivalently written as a SDP and solved by a generic interior-point solver~\cite{steffensCompactFormulationEll2018}. Before presenting the SDP reformulation, we introduce a compression of the measurement matrix $\vect{Y}$ in the oversampled case, i.e., $N > M$~\cite{liuMaximumPosterioriDirectionofarrival2024,yangSparseMethodsDirectionofarrival2018}.
Define
\begin{equation} \label{eq:measurementCompress}
  \widehat{\vect{Y}} = \vect{Y} \text{ if } N \leq M \text{ and } \widehat{\vect{Y}} = \sqrt{M} \cdot \widehat{\vect{R}}^{\frac{1}{2}} \text{ otherwise}.
\end{equation}
The compressed measurement matrix $\widehat{\vect{Y}}$ shares the sample covariance matrix $\widehat{\vect{R}}$ with the original measurement matrix $\vect{Y}$, i.e.,
$\widehat{\vect{Y}}\widehat{\vect{Y}}^\Herm / N' = \vect{Y}\vect{Y}^\Herm / N = \widehat{\vect{R}}$, whereas the number of columns of $\widehat{\vect{Y}}$, denoted by $N'$, is limited by $ \min \left\{M,N\right\} $. The objective function $f$ in~\eqref{prob:SI-SPARROW} can be rewritten as
\begin{equation} \label{eq:SI-SPARROW_obj_Y}
	f(\vect{Q}) = \tfrac{M}{N'} \tr \left( \widehat{\vect{Y}}^\Herm (\vect{Q} + \lambda \vect{I}_M)^{-1} \widehat{\vect{Y}}\right) + \tr (\vect{Q}).
\end{equation}
Then the following SDP reformulation for problem~\eqref{prob:SI-SPARROW}
\begin{equation} \label{prob:SI-SPARROW-SDP-N}
    \min_{{\vect{Q} \in \Sym_+^M \cap \set{T}^M, \, \vect{T} \in \Sym^{N'}}}  \quad  \tfrac{M}{N'}\tr (\vect{T}) + \tr (\vect{Q}) \qquad
		\st  \quad \begin{bmatrix}
			\vect{T} & \widehat{\vect{Y}}^{\Herm}\\ \widehat{\vect{Y}} & \vect{Q} + \lambda\vect{I}_{M}
		\end{bmatrix} \succeq 0 
\end{equation}
is obtained by using the reformulation technique in~\cite{steffensCompactFormulationEll2018}. 
With the compression in~\eqref{eq:measurementCompress}, the dimension of the semidefinite constraint in~\eqref{prob:SI-SPARROW-SDP-N}, which dominates the computational complexity of the SDP implementation~\eqref{prob:SI-SPARROW-SDP-N}, no longer grows with the number of snapshots $N$ in the oversampled case.
However, the above SDP implementation still becomes computationally intractable for a large number of sensors $ M $.
Hence, in Sections~\ref{sec:ADMM} and~\ref{sec:SCA}, we develop efficient iterative algorithms for the SI-SPARROW problem~\eqref{prob:SI-SPARROW} based on ADMM and SCA, each offering advantages in different scenarios.

\subsection{Connection to Other Methods}
\label{sec:otherMethods}

\subsubsection{GL-SPARROW and Atomic Norm Minimization (ANM)}
The shift-invariant structure~\eqref{eq:shift-invariant-subspace} is a generalization of the multilevel Toeplitz structure.
Hence, the SI-SPARROW problem~\eqref{prob:SI-SPARROW} is a relaxation of the GL-SPARROW problem. 
Whereas GL-SPARROW exploits a Toeplitz structure on $ \vect{Q} $ for ULAs, SI-SPARROW uses the shift-invariant structure~\eqref{eq:shift-invariant-subspace} to address the case with more general array geometries and calibration uncertainties.
Additionally, the GL-SPARROW method is shown to be equivalent to the widely-used ANM but with a reduced computational cost~\cite[Theorem 2]{steffensCompactFormulationEll2018}.
\subsubsection{2D-ANM}
As mentioned in Section~\ref{subsec:SPARROW}, as a generalization of ANM, the 2D-ANM method in~\cite{zhangEfficientTwodimensionalLine2019} uses a two-level Toeplitz structure to address the recovery of a 2D uniform sinusoid, e.g., the case with a 2D URA.

The relations among the above sets and their corresponding methods can be summarized in Fig.~\ref{fig:methodRelations}.
\begin{figure}
\begin{center}
  \begin{tikzpicture}[line width=1pt]
    \draw[fill=TUDa-0c] (0,0) circle [x radius=30mm, y radius=13mm];
    \node[text width=25mm] at (23mm,0) {\small Shift-invariant subspace $\set{T}^M$};
    \draw[fill=TUDa-0b] (-9mm,0) circle [x radius=18mm, y radius=10mm];
    \node[text width=20mm] at (3mm,0) {\small Multilevel Toeplitz matrices};
    \draw[fill=TUDa-0a] (-17mm,0) circle [x radius=8.5mm, y radius=5mm];
    \node[text width=20mm] at (-12mm,0) {\small Toeplitz matrices};
  \end{tikzpicture}
\begin{equation*}
  \text{GL-SPARROW} \equiv \text{ANM} \quad \xRightarrow{\text{generalization}} \quad \text{2D-ANM} \quad \xRightarrow{\text{generalization}} \quad \text{SI-SPARROW}
\end{equation*}
\end{center}
\caption{The relations among the various structures and their corresponding methods.}
\label{fig:methodRelations}
\end{figure}

The connections between GL-SPARROW and other similar gridless covariance-fitting methods that exploit the Toeplitz structure have been analyzed in~\cite{steffensCompactFormulationEll2018}, which are not repeated here since this work focuses on a generalization of Toeplitz structure that is applicable to more general array geometries.

As mentioned in Section~\ref{subsec:SPARROW}, the Carath\'eodory theorem~\cite{caratheodoryUberVariabilitatsbereichFourierschen1911,caratheodoryUberZusammenhangExtremen1911,toeplitzZurTheorieQuadratischen1911} ensures the uniqueness of the decomposition~\eqref{eq:dictConstraint} for any rank-deficient PSD Toeplitz matrix $\vect{Q}$ and its generalization to the multilevel Toeplitz case has been discussed in~\cite{yangVandermondeDecompositionMultilevel2016}. However, the conditions for the existence of a unique decomposition for the generalized shift-invariant structure~\eqref{eq:shift-invariant-subspace} are still unknown. 

\subsection{Alternative Expression for Shift-Invariant Subspace}
Before presenting the algorithms for solving~\eqref{prob:SI-SPARROW}, we introduce an alternative expression for the elements of the shift-invariant subspace $ \set{T}^M $.
The structural constraints~\eqref{eq:shift-invariance-constraint} enforce redundancy by requiring certain submatrices of $ \vect{Q} $ to be identical.
This means that any matrix $ \vect{Q} \in \set{T}^M $ can be equivalently expressed by a reduced set of independent variables.
For a matrix $ \vect{Q} \in \set{T}^M$, let $ \vect{q} = [q_1,\ldots,q_I]^\Trans \in \Compl^I $ be a vector that contains all the $ I $ independent variables in the upper triangle of $ \vect{Q} $ such that each entry of $ \vect{Q} $ contains either one variable in $ \vect{q} $ or its complex conjugate. In particular, let $ q_1 $ be the independent variable on the main diagonal of $ \vect{Q} $, which is real-valued since $ \vect{Q} $ is Hermitian.
Furthermore, define
\begin{equation}\label{eq:indMat}
	\vect{\Omega}_i = \tfrac{\partial \vect{Q}}{\partial q_i} \in \{0,1\}^{M \times M} \quad \text{for } i=1,\ldots, I,
\end{equation}
as the indicator matrix for the independent variable $ q_i $. 
The matrix $ \vect{\Omega}_i $ has a one in the entries of $ \vect{Q} $ occupied by $ q_i $ and a zero elsewhere. The entries of $ \vect{Q} $ occupied by the complex conjugate of $ q_i $ are then indicated by $ \vect{\Omega}_i^\Trans $ for $ i=2,\ldots,I $.
In summary, any matrix $ \vect{Q} \in \set{T}^M $ can be expressed as
\begin{equation}\label{eq:Q-structure}
	\begin{matrix}
		\vect{Q} = q_1 \vect{\Omega}_1 + \sum_{i=2}^{I} \big(q_i \vect{\Omega}_i + q_i^* \vect{\Omega}_i^\Trans\big)
	\end{matrix}
\end{equation}
with $q_1 \in \Real$ and $q_i \in \Compl$ for $i=2,\ldots,I$. Let $T(\cdot)$ denote the linear mapping in~\eqref{eq:Q-structure} from the vector $\vect{q}$ to the matrix $\vect{Q}$.
Based on the expression in~\eqref{eq:Q-structure}, the function $ f $, in the subspace $ \set{T}^M $, can be regarded as a function dependent only on $ \vect{q} $. Hence, the notations $ f(\vect{Q}) $ and $ f(\vect{q}) $ are used interchangeably in this paper when only elements of $ \set{T}^M $ are considered.

\subsection{Multi-Invariance Multidimensional ESPRIT}
\label{sec:MI-MD-ESPRIT}


A limitation of MD-ESPRIT is that it can only consider a single shift-invariance equation per dimension. To exploit all possible intrasubarray shift invariances, a simple extension of MD-ESPRIT, termed multi-invariance multidimensional ESPRIT (MI-MD-ESPRIT), is proposed in this section.

Moreover, the shift invariances in~\eqref{eq:shift-invariance_sensor_x} and~\eqref{eq:shift-invariance_sensor_y} imply that a complete list of intrasubarray shift invariances is given by 
\begin{subequations} \label{eq:shift-invariance_intrasubarray_all}
  \begin{align}
    {(\vect{K}_{l}^{x})}^\Trans \vect{A}(\bmu,\vect{\alpha}) & = {(\vect{K}_{k}^{x})}^\Trans   \vect{A} (\bmu,\vect{\alpha})  \vect{\Phi}\left((\delta_{l}^{x} - \delta_k^x)\bmu^x \right) 
                                                             & \text{for } k,l = 1,\dots,L_x \text{ and } k < l, \label{eq:shift-invariance_sensor_x_all} \\
    {(\vect{K}_{l}^{y})}^\Trans \vect{A}(\bmu,\vect{\alpha}) & = {(\vect{K}_k^{y})}^\Trans   \vect{A} (\bmu,\vect{\alpha})  \vect{\Phi}\left((\delta_{l}^{y} - \delta_k^y)\bmu^y \right)  
                                                             & \text{for } k, l = 1,\dots,L_y \text{ and } k < l. \label{eq:shift-invariance_sensor_y_all}
  \end{align}
\end{subequations}
Other than those in~\eqref{eq:shift-invariance_sensor_x} and~\eqref{eq:shift-invariance_sensor_y}, the remaining shift-invariance equations in~\eqref{eq:shift-invariance_intrasubarray_all} are redundant for the construction of the shift-invariant structure~\eqref{eq:shift-invariance-constraint}. However, due to the finite sample size and the effect of noise in practical applications, the estimation accuracy can be further improved when additional shift invariances are included in an ESPRIT-like method.

To utilize all the intrasubarray shift invariances in~\eqref{eq:shift-invariance_intrasubarray_all} that involve the known sensor displacements $\delta_l^x$ for $l = 2, \ldots, L_x$ and $\delta_l^y$ for $l = 2, \ldots, L_y$, we extend the MD-ESPRIT method by performing the following two steps.
Define
\begin{equation} \label{eq:shiftDiagMat}
	\begin{aligned}
    \vect{\Phi}_{k,l}^x & \! = \! \vect{\Phi} \left((\delta_l^x \! - \! \delta_k^x) \vect{\mu}^x \right) \ \text{for } k,l \! = \! 1,\ldots,L_x \text{ and } k \! < \! l, \\
    \vect{\Phi}_{k,l}^y & \! = \! \vect{\Phi} \left((\delta_l^y \! - \! \delta_k^y) \vect{\mu}^y \right) \ \text{for } k,l \! = \! 1,\ldots,L_y \text{ and } k \! < \! l.
	\end{aligned}
\end{equation}
First, to separate the sources, MD-ESPRIT is performed on an augmented problem with the above coupling of the diagonal matrices $\vect{\Phi}_{k,l}^x$ or $\vect{\Phi}_{k,l}^y$ in~\eqref{eq:shiftDiagMat} for the same dimension $ x $ or $ y $, respectively, ignored.
Specifically, the diagonal matrices $\vect{\Phi}_{k,l}^x$ and $\vect{\Phi}_{k,l}^y$ are recovered by performing the simultaneous diagonalization on the shift-invariance equations in~\eqref{eq:shift-invariance_intrasubarray_all}.
Then, the relation in~\eqref{eq:shiftDiagMat} is used to further recover the frequencies as follows.
The entries of the recovered diagonal matrices in~\eqref{eq:shiftDiagMat} are used to reconstruct augmented 1D subarray steering vectors for each frequency to be estimated defined as
\begin{equation} \label{eq:subarraySteerVec_augmented}
	\begin{aligned}
    \widetilde{\vect{v}}^x (\mu_i^x) &= \big[1, \euler^{\imagunit \mu_i^x \delta^x_2}, \ldots, \euler^{\imagunit \mu_i^x \delta^x_{L_x}}, \euler^{\imagunit \mu_i^x (\delta_3^x - \delta_2^x)}, 
    \ldots, \euler^{\imagunit \mu_i^x (\delta_{L_x}^x - \delta_2^x)}, \ldots, \euler^{\imagunit \mu_i^x (\delta_{L_x}^x - \delta_{L_x-1}^x)}\big]^\Trans, \\
    \widetilde{\vect{v}}^y (\mu_i^y) &= \big[1, \euler^{\imagunit \mu_i^y \delta^y_2}, \ldots, \euler^{\imagunit \mu_i^y \delta^y_{L_y}}, \euler^{\imagunit \mu_i^y (\delta_3^y - \delta_2^y)}, 
    \ldots, \euler^{\imagunit \mu_i^y (\delta_{L_y}^y - \delta_2^y)}, \ldots, \euler^{\imagunit \mu_i^y (\delta_{L_y}^y - \delta_{L_y-1}^y)}\big]^\Trans
	\end{aligned}
\end{equation}
for $i=1,\ldots,N_\text{S}$.
With the known intrasubarray displacements $\delta_l^x$ and $\delta_l^y$, the recovery of the frequencies can be performed independently from each reconstructed augmented 1D steering vector in~\eqref{eq:subarraySteerVec_augmented} by a single-source recovery method, e.g., the deterministic maximum likelihood (DML) estimator via a simple 1D grid search. 

\section{ADMM for Shift-Invariant SPARROW}
\label{sec:ADMM}

\begin{algorithm}[t]
	\caption{ADMM Algorithm for SI-SPARROW.}
	\label{alg:ADMM}
  \KwIn{$\widehat{\vect{R}}$, $ \{\vect{\Omega}_i\}_{i=1}^I $, $\lambda > 0$
  }
		Initialize the variables according to~\eqref{eq:ADMM_initialization} and $t \gets 0$\;
		\While{not converged}{
			Update $ \vect{Q}^{(t)} $ by solving~\eqref{prob:Q-update} using Algorithm~\ref{alg:innerSCA}\;
			$ \vect{Z}^{(t+1)} \gets \opr{P}_{\Sym_+^M} \left( \vect{Q}^{(t+1)} + \vect{U}^{(t)} \right) $\;
			$\vect{U}^{(t+1)} \gets \vect{U}^{(t)} + \vect{Q}^{(t+1)} - \vect{Z}^{(t+1)}$\;
			$t \gets t+1$\;
			
		}
		\Return{$\vect{Q}^{(t)}$\;}

\end{algorithm}

In this section, we use the ADMM framework to solve the SI-SPARROW problem~\eqref{prob:SI-SPARROW}. The choice of ADMM is motivated by the fact that problem~\eqref{prob:SI-SPARROW} can be easily solved when only one of the two kinds of constraints, i.e., either the PSD constraint or the shift-invariance constraints, is enforced. The ADMM framework is used to decompose the problem such that, in each subproblem, only one of the constraints needs to be fulfilled. We assume that the sample covariance matrix $ \widehat{\vect{R}} $ is positive definite, which is required to ensure the convergence of the solution approach for the primal subproblem in ADMM introduced later. In the undersampled case, i.e., $ N < M $, where the sample covariance matrix is always rank deficient, a small positive diagonal loading onto the sample covariance matrix is performed before solving~\eqref{prob:SI-SPARROW}.

To apply ADMM, we first reformulate problem~\eqref{prob:SI-SPARROW} as
\begin{equation} \label{prob:ADMM}
    \underset{{\vect{Q} \in \mathcal{T}^M, \, \vect{Z} \in \Sym^M} }{\min}  \quad f(\vect{Q}) + g(\vect{Z}) \qquad
    \text{s.t.} \quad \vect{Q} - \vect{Z} = \vect{0}, 
		\ \vect{Q} + \lambda \vect{I}_M \succ 0, 
\end{equation}
where $g(\vect{Z}) = \mathbb{I}_{\Sym_+^M} (\vect{Z})$.
In~\eqref{prob:ADMM}, by introducing an auxiliary variable $ \vect{Z} $, the shift-invariance constraints~\eqref{eq:shift-invariance-constraint} are separated from the PSD constraint $ \vect{Q} \succeq 0 $ so that the two types of constraints can be addressed alternately.
As a necessary condition of $ \vect{Q} = \vect{Z} \succeq 0 $, the constraint $ \vect{Q} + \lambda \vect{I}_M \succ 0 $ is redundant. 
However, it ensures the convexity of the primal subproblem in ADMM that is introduced in~\eqref{prob:Q-update}. 

The augmented Lagrangian of problem~\eqref{prob:ADMM} is~\cite{boydDistributedOptimizationStatistical2011}
\begin{equation}
	L_{\rho} (\vect{Q},\vect{Z},\vect{U}) = f(\vect{Q}) + g(\vect{Z}) + \tfrac{\rho}{2} \| \vect{Q} - \vect{Z} + \vect{U} \|_\Frob^2,
\end{equation}
where $ \rho > 0 $ is the penalty weight and $ \vect{U} \in \Sym^M $ is the dual variable scaled by $ 1/ \rho $. 
Let \(\vect{Q}^{(t)}, \vect{Z}^{(t)}, \vect{U}^{(t)}\) be the values of the primal and dual variables at iteration $t$.
The scaled form of ADMM for~\eqref{prob:ADMM} has the following steps in each iteration:
\begin{subequations} \label{eq:ADMMiter}
	\begin{align}
    \vect{Q}^{(t+1)} &= \underset{\vect{Q} \in \mathcal{T}^M}{\argmin} \ L_\rho (\vect{Q},\vect{Z}^{(t)},\vect{U}^{(t)}) \quad \text{s.t.} \quad \vect{Q} + \lambda \vect{I}_M \succ 0, \label{prob:Q-update} \\
		\vect{Z}^{(t+1)} &= \underset{\vect{Z} \in \Sym^M}{\argmin} \ L_\rho (\vect{Q}^{(t+1)},\vect{Z},\vect{U}^{(t)}) \nonumber \\
    &= \opr{P}_{\Sym_+^M} \big( \vect{Q}^{(t+1)} + \vect{U}^{(t)} \big), \label{prob:Z-update} \\
		\vect{U}^{(t+1)} &= \vect{U}^{(t)} + \vect{Q}^{(t+1)} - \vect{Z}^{(t+1)}.
	\end{align}
\end{subequations}
The projection onto the PSD cone
\(\Sym_+^M\), denoted by \(\opr{P}_{\Sym_+^M}\), is obtained by discarding the components with negative eigenvalues in the eigenvalue decomposition (EVD) of the argument~\cite{highamComputingNearestSymmetric1988}.
The subproblem for the \(\vect{Q}\)-update~\eqref{prob:Q-update} is convex. Define \(\bar{\vect{Q}}^{(t)} = \vect{Z}^{(t)} - \vect{U}^{(t)}\). The subproblem~\eqref{prob:Q-update} can be viewed as a proximal mapping at $\bar{\vect{Q}}^{(t)}$, which, in contrast, has no closed-form solution. Hence, in Section~\ref{subsec:Q-update}, we develop an efficient iterative algorithm for solving subproblem~\eqref{prob:Q-update} based on the SCA framework with separable quadratic approximation.

The ADMM algorithm for problem~\eqref{prob:SI-SPARROW} is outlined in Algorithm~\ref{alg:ADMM}, including the initialization strategy that will be discussed in Section~\ref{sec:ADMM_initialization}.
The global convergence of ADMM is established under the assumptions in~\cite[Sec. 3.2]{boydDistributedOptimizationStatistical2011},
which can be readily verified for the considered convex problem~\eqref{prob:ADMM}. 

\subsection{Successive Convex Approximation for the \texorpdfstring{$\vect{Q}$}{Q}-update}
\label{subsec:Q-update}

\begin{algorithm}[t]
	\caption{SCA Algorithm for the $ \vect{Q} $-update~\eqref{prob:Q-update}.}
	\label{alg:innerSCA}
  \KwIn{$\bar{\vect{Q}}$, $ \{\vect{\Omega}_i\}_{i=1}^I $, $\lambda>0$}
	Initialize $\vect{Q}^{(0)}$ and $l \gets 0$\;
	\While{not converged}{
		Compute $ \widetilde{\vect{q}}^{(l)} $ according to~\eqref{eq:approxSol_1} and~\eqref{eq:approxSol_2}\;
		$\widetilde{\vect{Q}}^{(l)} \gets T(\widetilde{\vect{q}}^{(l)})$\;
		Compute step size $ \gamma^{(l)} $ by Armijo rule described in Section~\ref{subsubsec:stepsize}\;
		$ \vect{Q}^{(l+1)} \gets \vect{Q}^{(l)} + \gamma^{(l)} (\widetilde{\vect{Q}}^{(l)} - \vect{Q}^{(l)}) $\;
		$l \gets l+1$\;
	}
	\Return $\vect{Q}^{(l)}$\;
	
\end{algorithm}

Since the subproblem for the $ \vect{Q} $-update~\eqref{prob:Q-update} is convex, a global optimum can be efficiently obtained by using the SCA framework~\cite{yangUnifiedSuccessivePseudoconvex2017} with separable quadratic approximation.
As we discuss only a single ADMM iteration, for simplicity, the ADMM iteration index $ t $ in $ \bar{\vect{Q}}^{(t)} $ is omitted and the objective function in~\eqref{prob:Q-update} is denoted by
\begin{equation}
	h(\vect{Q}) = f(\vect{Q}) + \tfrac{\rho}{2} \norm{\vect{Q} - \bar{\vect{Q}}}_\Frob^2,
\end{equation}
where $g(\vect{Z})$ as a constant of $\vect{Q}$ is also omitted.
In each iteration, we first construct a separable quadratic approximation for \(h\) by exploiting the shift-invariant structure of $ \vect{Q} $.
To obtain an approximate problem that can be decomposed and solved in parallel, we minimize the quadratic approximate function in a relaxed domain of problem~\eqref{prob:Q-update} with the positive definiteness constraint discarded. The solution of the quadratic approximate problem indicates a descent direction of the original function $ h $.
Then, $\vect{Q}$ is updated along this direction using a step size determined by an Armijo-rule-based backtracking line search to ensure sufficient decrease and feasibility. 

When $ \widehat{\vect{R}} \succ 0 $, the term $ \tr \big((\vect{Q} + \lambda \vect{I}_M)^{-1} \widehat{\vect{R}}\big) $ in $ f $ acts as a barrier for the set $ \{\vect{Q} \in \Sym^M \mid \vect{Q} + \lambda \vect{I}_M 
\succ 0\} $, i.e., both its function value and gradient tend to $ + \infty $ when $ \vect{Q} $ approaches the boundary of the set from the interior. Thus, the exclusion of the positive definiteness constraint in~\eqref{prob:Q-update} from the approximate problem does not impact the convergence of the proposed algorithm. It converges to a global optimum of the convex problem~\eqref{prob:Q-update} by the regularity conditions of the SCA framework in~\cite{yangUnifiedSuccessivePseudoconvex2017}.
In the undersampled case, i.e., $N < M$, where $\widehat{\vect{R}}$ is rank deficient, a small diagonal loading is added to ensure convergence.
Moreover, the proposed algorithm can also be viewed as a variant of the diagonal approximation to Newton's method~\cite{bertsekasNonlinearProgramming2016}, 
which is shown to have an asymptotically quadratic convergence rate.

\subsubsection{Separable Quadratic Approximation}

Using the expression in~\eqref{eq:Q-structure}, we design a convex quadratic approximation for the $ \vect{Q} $-update subproblem~\eqref{prob:Q-update} that is separable in the independent variables $ \vect{q} $.
To this end, we first present the expressions for the gradient and quadratic Taylor expansion of $ f $. When only the Hermitian structure of $ \vect{Q} $, and not the shift-invariance structural constraints, is considered, the gradient of \(f\) in~\eqref{prob:ADMM} with respect to \(\vect{Q}\)
is given by
\begin{equation}\label{eq:gradf_Q}
	\nabla_\vect{Q} f(\vect{Q}) = - M (\vect{Q} + \lambda \vect{I}_M)^{-1} \widehat{\vect{R}} (\vect{Q} + \lambda \vect{I}_M)^{-1} + \vect{I}_M,
\end{equation}
and the gradient of $ h $ with respect to $ \vect{Q} $ is
\begin{equation}\label{eq:gradh_Q}
	\nabla_{\vect{Q}} h(\vect{Q}) = \nabla_{\vect{Q}} f (\vect{Q}) + \rho \big( \vect{Q} - \bar{\vect{Q}}\big).
\end{equation}
Let $ \vect{Q}^{(l)} $ be the approximate solution for problem~\eqref{prob:Q-update} in the $ l $th iteration and $ \vect{q}^{(l)} $ be the corresponding values for the independent variables $ \vect{q} $. 
The quadratic Taylor expansion of $ f $ around $ \vect{Q}^{(l)} $ can then be expressed as
\begin{equation}\label{eq:fTaylor}
	f(\vect{Q}) \approx f (\vect{Q}^{(l)}) + \tr \big(\nabla_{\vect{Q}} f(\vect{Q}^{(l)}) \vect{\Delta}\big) + M \tr \big(\vect{V \Delta} \widetilde{\vect{R}} \vect{\Delta}\big),
\end{equation}
where $ \vect{\Delta} = \vect{Q} - \vect{Q}^{(l)} $, $ \vect{V} = (\vect{Q}^{(l)} + \lambda \vect{I}_M)^{-1} $, and $ \widetilde{\vect{R}} = \vect{V} \widehat{\vect{R}} \vect{V} $.
\iflong
The details of the computation of the gradient and second-order Taylor expansion of $f$ are provided in Appendix~\ref{appendix:Taylor}.
\fi

Departing from Newton's method, which iteratively approximates the original function by its second-order Taylor expansion, 
we adopt a separable quadratic approximation,
which preserves only the diagonal entries of the augmented Hessian of $ h $ with respect to the independent variables $ \vect{q} $ at $ \vect{q}^{(l)} $.
In this case, instead of using the augmented form of gradient and Hessian~\cite{sayedAdaptationLearningOptimization2014}, the separable quadratic approximation can be expressed with the following compact form of partial gradient and Hessian of $ h $ with respect to a complex variable $ q_i $:
\begin{equation} \label{eq:partialGradient}
  \nabla_{q_i} h (\vect{q}) = 2 \tfrac{\partial}{\partial q_i^*} h(\vect{q}) \quad \text{and} \quad
  \nabla_{q_i}^2 h (\vect{q}) = 2 \tfrac{\partial^2}{\partial q_i^* \partial q_i} h(\vect{q}),
\end{equation}
where $ \tfrac{\partial}{\partial q_i} $ and $ \tfrac{\partial}{\partial q_i^*} $ are the Wirtinger derivative operators~\cite{sayedAdaptationLearningOptimization2014}.
With the above notation, the separable quadratic approximation of $ h $ around $ \vect{q}^{(l)} $ is given by
\begin{align}
	\widetilde{h}^{(l)} (\vect{q}) &= h(\vect{q}^{(l)}) + \begin{matrix}
		\sum_{i = 1}^I
	\end{matrix} \Re \big( \nabla_{q_i}^* h(\vect{q}^{(l)}) ( q_i - q_i^{(l)} ) \big) 
	+ \begin{matrix}
		\sum_{i = 1}^I
	\end{matrix} \tfrac{1}{2} \nabla_{q_i}^2 h(\vect{q}^{(l)}) \big| q_i - q_i^{(l)} \big|^2. \label{eq:h_approx}
\end{align}
From the results in~\eqref{eq:gradh_Q} and~\eqref{eq:fTaylor}, the partial gradient and Hessian of $ h $ with respect to an independent variable $ q_i $ can be calculated by the chain rule of differentiation as 
\begin{subequations}
	\begin{align}
    \nabla_{q_i} h(\vect{q}^{(l)}) &= 2 \tr \big(\vect{\Omega}_i^\Trans \cdot \nabla_{\vect{Q}} h(\vect{Q}^{(l)}) \big), \label{eq:partialGradh} \\
    \nabla_{q_i}^2 h(\vect{q}^{(l)}) &=  2 M \tr \big(\widetilde{\vect{R}} (\vect{\Omega}_i^\Trans \vect{V} \vect{\Omega}_i + \vect{\Omega}_i \vect{V \Omega}_i^\Trans )\big) 
    + 2 \rho \|\vect{\Omega}_i\|_\Frob^2 \label{eq:partialHessh}
	\end{align}
\end{subequations}
for $ i \!=\! 2,\ldots,I $, where the matrix $\vect{\Omega}_i$ is defined in~\eqref{eq:indMat} and indicates the entries of $\vect{Q}$ occupied by the independent variable $q_i$.
Recall that all diagonal elements of the matrix $\vect{Q}$ are identical and real-valued, which are represented by the independent variable $q_1$.
For the real-valued variable $ q_1 $ the factor of two in~\eqref{eq:partialHessh} and~\eqref{eq:partialGradh} needs to be removed.

Then the following separable approximate problem is constructed at the $ l $th iteration:
\begin{equation}\label{prob:Q-update-approx}
	\widetilde{\vect{q}}^{(l)} = {\argmin}_{\vect{q} \in \Compl^I} \quad \widetilde{h}^{(l)} (\vect{q}) \quad \st \quad q_1 \geq -\lambda,
\end{equation}
where, as aforementioned, the positive definiteness constraint in~\eqref{prob:Q-update} is relaxed to a trivial separable bounding constraint.
Problem~\eqref{prob:Q-update-approx} can be decomposed into $ I $ independent subproblems. Each subproblem depends exclusively on a single variable in $ \vect{q} $ and, hence, can be solved in parallel. Specifically, each subproblem is a univariate convex quadratic program, which admits the following closed-form solution:
\begin{equation}\label{eq:approxSol_1}
  \widetilde{q}_i^{(l)} = q_i^{(l)} - {\nabla_{q_i} h(\vect{q}^{(l)})} / {\nabla_{q_i}^2 h(\vect{q}^{(l)})} \quad \text{for } i = 2,\ldots,I.
\end{equation}
With the bounding constraint in~\eqref{prob:Q-update-approx}, the solution of the subproblem involving $q_1$ is given by
\begin{equation}\label{eq:approxSol_2}
	\widetilde{q}_1^{(l)} = \max \left\{ -\lambda, q_1^{(l)} - {\nabla_{q_1} h(\vect{q}^{(l)})} / {\nabla_{q_1}^2 h(\vect{q}^{(l)})} \right\}.
\end{equation}

The matrix $ \widetilde{\vect{Q}}^{(l)} $ corresponding to $ \widetilde{\vect{q}}^{(l)} $ is then computed according to~\eqref{eq:Q-structure}.
Define \(\vect{\Delta}^{(l)} = \widetilde{\vect{Q}}^{(l)} - \vect{Q}^{(l)}\).
Although $ \widetilde{\vect{Q}}^{(l)} $ may not satisfy the positive definiteness constraint in~\eqref{prob:Q-update},
\(\vect{\Delta}^{(l)}\) shows a descent direction of the original function $ h $ due to the gradient consistency between $ h $ and the approximation $ \widetilde{h}^{(l)} $ at the current point \(\vect{Q}^{(l)}\). 
To ensure feasibility, as well as sufficient decrease of the original function $ h $, we perform a line search along the descent direction $ \vect{\Delta}^{(l)} $, updating
\begin{equation}\label{eq:lineSearch}
	\vect{Q}^{(l+1)} = \vect{Q}^{(l)} + \gamma^{(l)} \vect{\Delta}^{(l)}
\end{equation}
with a suitable step size \(\gamma^{(l)} \in (0,1]\).
Moreover, when $ \widetilde{\vect{Q}}^{(l)} = \vect{Q}^{(l)} $, a stationary point of $ \widetilde{h}^{(l)} $ is achieved, which is also a stationary point and global optimum of the original function $ h $ due to the gradient consistency~\cite[Thm. 1]{yangUnifiedSuccessivePseudoconvex2017}.

\subsubsection{Step Size Computation for the \texorpdfstring{$\vect{Q}$}{Q}-update subproblem}
\label{subsubsec:stepsize}

The Armijo-rule-based backtracking line search is employed to find a suitable step size for the update~\eqref{eq:lineSearch} at a low computational cost. In the Armijo rule, step sizes \(\gamma \in \{ \beta^0, \beta^1, \ldots\}\), i.e., a geometric sequence with a decrease rate \(0 < \beta < 1\), are successively tried until the smallest \(k \in \mathbb{N}\) with \(\gamma = \beta^k\) is found such that \(\vect{Q}^{(l)} + \gamma \vect{\Delta}^{(l)}\) is feasible and satisfies the following sufficient decrease condition:
\begin{equation}\label{eq:armijo}
	h \big( \vect{Q}^{(l)} + \gamma \vect{\Delta}^{(l)} \big) \leq h(\vect{Q}^{(l)}) + \gamma \sigma \text{tr} \big(\nabla h(\vect{Q}^{(l)})^\Herm \vect{\Delta}^{(l)}\big),
\end{equation}
where \(0 < \sigma <1\). 
The positive definiteness can be verified while we compute the Cholesky decomposition that is required for the matrix inversion in the objective function evaluation. The computational cost in the step size computation is dominated by the Cholesky decomposition (costs $ M^3/3 $ flops).

Finally, the proposed SCA-based algorithm for the $ \vect{Q} $-update subproblem~\eqref{prob:Q-update} is outlined in Algorithm~\ref{alg:innerSCA}. As for the initialization, the variable $ \vect{Q} $ is simply initialized to be the current value \(\vect{Q}^{(t)}\) in the ADMM iteration.


\subsection{Initialization}
\label{sec:ADMM_initialization}
Provided that the sample covariance matrix $ \widehat{\vect{R}} $ is positive definite, the objective function $ f(\vect{Q}) = M \tr \big(({\vect{Q}} + \lambda \vect{I}_{M})^{-1}\widehat{\vect{R}} \big)+\tr (\vect{Q}) $ in~\eqref{prob:SI-SPARROW} is convex in the set $ \{\vect{Q} \in \Sym^M \mid \vect{Q} + \lambda \vect{I}_M \succ 0\} $ and, hence, has a unique minimum point, which can be obtained by solving the stationarity condition
\begin{equation}\label{eq:stationarity}
	\nabla_\vect{Q} f(\vect{Q}) = \vect{0} \quad \text{for} \quad \vect{Q} + \lambda \vect{I}_M \succ 0
\end{equation}
with $ \nabla_{\vect{Q}} f $ given in~\eqref{eq:gradf_Q}.
That is, the relaxed problem
\begin{equation}\label{prob:relaxed1}
	\widehat{\vect{Q}}_1 = {\argmin}_{\vect{Q} \in \Sym^M} \ f(\vect{Q}) \quad
	\text{s.t.} \quad \vect{Q} + \lambda \vect{I}_M \succ 0
\end{equation}
admits the closed-form solution
	$\widehat{\vect{Q}}_1 = \sqrt{M} \widehat{\vect{R}}^{\frac{1}{2}} - \lambda \vect{I}_M$,
which is obtained from the stationarity condition in~\eqref{eq:stationarity}.

Due to the noise in the measurements, \(\widehat{\vect{Q}}_1\) is not guaranteed to be PSD and satisfy the shift-invariance constraints. Therefore, we then solve the following relaxation to~\eqref{prob:ADMM}:
\begin{equation}\label{prob:ADMM_relax}
	\widehat{\vect{Q}}_2 = {\argmin}_{\vect{Q} \in \mathcal{T}^M} \ f(\vect{Q}) \quad
	\text{s.t.} \quad \vect{Q} + \lambda \vect{I}_M \succ 0,
\end{equation}
where the PSD constraint on $ \vect{Q} $ is discarded. The relaxed problem~\eqref{prob:ADMM_relax} can be viewed as an instance of the \(\vect{Q}\)-update subproblem~\eqref{prob:Q-update} with the penalty parameter \(\rho=0\).
Hence, problem~\eqref{prob:ADMM_relax} can be solved by Algorithm~\ref{alg:innerSCA} with an initialization $ \vect{Q}^{(0)} = \opr{P}_{\set{T}^M} (\widehat{\vect{Q}}_1) + \iota \vect{I}_M $, where $ \iota \geq 0 $ is chosen to ensure the positive definiteness of $ \vect{Q}^{(0)} + \lambda \vect{I}_M $.  
The projection onto set \(\mathcal{T}^M\) is easily implemented by averaging the entries corresponding to the same independent variable.

If the solution \(\widehat{\vect{Q}}_2\) of~\eqref{prob:ADMM_relax} is PSD, then it is also the optimal solution of problem~\eqref{prob:ADMM} and there is no need to perform the ADMM iterations in~\eqref{eq:ADMMiter}. 
Otherwise, the primal and dual variables for the ADMM iterations are initialized as
\begin{equation} \label{eq:ADMM_initialization}
		\vect{Q}^{(0)} = \widehat{\vect{Q}}_2, \quad
		\vect{Z}^{(0)} = \opr{P}_{\Sym_+^M} ( \vect{Q}^{(0)}), \quad
		\vect{U}^{(0)} = \vect{Q}^{(0)} - \vect{Z}^{(0)}.
\end{equation}

\begin{algorithm}[t]
	\caption{SCA Algorithm for SI-SPARROW.}
	\label{alg:SCA}
  \KwIn{$\widehat{\vect{R}}$, $ \{\vect{\Omega}_i\}_{i=1}^I $, $\lambda>0$}
		Initialize $\vect{Q}$ according to~\eqref{eq:SCA_initialization} and $t \gets 0$\;
		\While{not converged}{
			Compute $ \widetilde{\vect{Q}}^{(t)} $ by solving the approximate problem~\eqref{prob:quadraticApprox} using ADMM algorithm\;
			Compute step size $ \gamma^{(t)} $ by Armijo rule\;
			$ \vect{Q}^{(t+1)} \gets \vect{Q}^{(t)} + \gamma^{(t)} (\widetilde{\vect{Q}}^{(t)} - \vect{Q}^{(t)}) $\;
			$t \gets t+1$\;
			
		}
		\Return{$\vect{Q}^{(t)}$\;}
	
\end{algorithm}

\section{Successive Convex Approximation for Shift-Invariant SPARROW}
\label{sec:SCA}


As discussed in Section~\ref{sec:ADMM}, the convergence of the ADMM Algorithm~\ref{alg:ADMM} requires the sample covariance matrix $\widehat{\vect{R}}$ to be positive definite. In the undersampled case, this is enforced by a small positive diagonal loading onto $\widehat{\vect{R}}$, which may result in a degradation of the estimation quality. On the other hand, the SDP approach based on the reformulation in~\eqref{prob:SI-SPARROW-SDP-N} relies on an efficient SDP solver, which often has high hardware requirements and implementation costs.
Thus, in this section, by switching the inner and outer iterations in Algorithm~\ref{alg:ADMM}, we develop an alternative algorithm for the SI-SPARROW problem~\eqref{prob:SI-SPARROW}, ensuring convergence without requiring the positive definiteness of the sample covariance matrix and enabling implementation on relatively low-cost hardware. In particular, we apply the SCA framework directly on the SI-SPARROW problem~\eqref{prob:SI-SPARROW}, where a descent direction of the function $f$ in~\eqref{prob:SI-SPARROW} in the set $\Sym_+^M \cap \set{T}^M$ is obtained by solving a quadratic approximate problem and then the backtracking line search is employed to update the variable $\vect{Q}$ along the descent direction. Nevertheless, unlike Algorithm~\ref{alg:innerSCA} for the $\vect{Q}$-update subproblem in~\eqref{prob:Q-update}, the quadratic approximation for the original problem~\eqref{prob:SI-SPARROW} cannot be solved in closed form since $\vect{Q}$ is required to be in both the PSD cone $\Sym_+^M$ and the shift-invariant subspace $\set{T}^M$. Hence, the ADMM framework is used to iteratively solve the quadratic approximation in an inner loop by separating the PSD constraint from the shift-invariance constraints.

Similar to the approximation in~\eqref{eq:h_approx}, to obtain an approximate problem that can be solved in parallel at a low computational cost, we construct the separable quadratic approximation
\begin{align}
	\widetilde{f}^{(t)} (\vect{q}) &= f(\vect{q}^{(t)}) + \begin{matrix}
		\sum_{i = 1}^I
	\end{matrix} \Re \big( \nabla_{q_i}^* f(\vect{q}^{(t)}) ( q_i - q_i^{(t)} ) \big) 
	+ \begin{matrix}
		\sum_{i = 1}^I
	\end{matrix} \tfrac{1}{2} \nabla_{q_i}^2 f(\vect{q}^{(t)}) \big| q_i - q_i^{(t)} \big|^2, \label{eq:f_approx}
\end{align}
which preserves only the diagonal entries of the Hessian of the original function $ f $ in~\eqref{prob:SI-SPARROW} with respect to the independent variables $ \vect{q} $ at the current point $ \vect{q}^{(t)} $.
In~\eqref{eq:f_approx}, the partial Hessian and gradient of $ f $ with respect to an independent variable $ q_i $ can likewise be calculated by the chain rule of differentiation from the results in~\eqref{eq:gradf_Q} and~\eqref{eq:fTaylor}. They are expressed as 
\begin{subequations}
	\begin{align}
		\nabla_{q_i}^2 f(\vect{q}^{(t)}) &= 2M \tr \big(\widetilde{\vect{R}} (\vect{\Omega}_i^\Trans \vect{V} \vect{\Omega}_i + \vect{\Omega}_i \vect{V \Omega}_i^\Trans )\big), \label{eq:partialHessf} \\
		\nabla_{q_i} f(\vect{q}^{(t)}) &= 2 \tr \big(\vect{\Omega}_i^\Trans \cdot \nabla_{\vect{Q}} f(\vect{Q}^{(t)}) \big), \label{eq:partialGradf}
	\end{align}
\end{subequations}
with $ \vect{V} = (\vect{Q}^{(t)} + \lambda \vect{I}_M)^{-1} $ and $ \widetilde{\vect{R}} = \vect{V} \widehat{\vect{R}} \vect{V} $ for $i=2,\ldots,I$. For the real-valued variable $ q_1 $ on the main diagonal of matrix $ \vect{Q} $, the factor of two in~\eqref{eq:partialHessf} and~\eqref{eq:partialGradf} needs to be removed. Then the quadratic approximation for the original SI-SPARROW problem~\eqref{prob:SI-SPARROW} at the $t$th iteration reads as
\begin{equation}\label{prob:quadraticApprox}
	\widetilde{\vect{Q}}^{(t)} = \argmin_{\vect{Q} \in \Sym_+^M \cap \set{T}^M} \ \widetilde{f}^{(t)} (\vect{Q}).
\end{equation}

Although $\widetilde{f}^{(t)}$ is separable in the independent variables $\vect{q}$, problem~\eqref{prob:quadraticApprox}, unfortunately, cannot be solved in closed form due to the PSD constraint. Due to the convexity of $f$, the approximate function $\widetilde{f}^{(t)}$ is a convex quadratic function. Thus, problem~\eqref{prob:quadraticApprox} can be viewed as a projection onto the intersection of two convex sets, i.e., the PSD cone $\Sym_+^M$ and the shift-invariant subspace $\set{T}^M$, which can be conventionally solved by the alternating projection~\cite{suffridgeApproximationHermitianPositive1993} that alternately projects onto each of the two sets. On the other hand, the ADMM framework, when applied to the projection problem in~\eqref{prob:quadraticApprox}, reduces to the Dykstra's alternating projection~\cite{boydDistributedOptimizationStatistical2011,dykstraAlgorithmRestrictedLeast1983}, which is shown to be far more efficient than the classical alternating projection that does not use a dual variable~\cite{bauschkeDykstrasAlternatingProjection1994}. Hence, we employ the ADMM algorithm to solve the approximate problem~\eqref{prob:quadraticApprox}, which, similar to~\eqref{eq:ADMMiter}, consists of the following steps in the $l$th inner-iteration:
\begin{subequations} \label{eq:ADMMiter_inner}
	\begin{align}
		\vect{Q}^{(l+1)} &= \underset{\vect{Q} \in \mathcal{T}^M}{\argmin} \ \widetilde{f}^{(t)}(\vect{Q}) + \tfrac{\rho}{2} \norm{\vect{Q} - \vect{Z}^{(l)} + \vect{U}^{(l)} }_\Frob^2, \label{prob:Q-update_inner} \\
		\vect{Z}^{(l+1)} &= \opr{P}_{\Sym_+^M} \big( \vect{Q}^{(l+1)} + \vect{U}^{(l)} \big), \\
		\vect{U}^{(l+1)} &= \vect{U}^{(l)} + \vect{Q}^{(l+1)} - \vect{Z}^{(l+1)}.
	\end{align}
\end{subequations}
Unlike~\eqref{eq:ADMMiter}, the $\vect{Q}$-update subproblem in~\eqref{prob:Q-update_inner} is an unconstrained quadratic program in the independent variables $\vect{q}$ and admits a closed-form solution. 
Moreover, since $ \widetilde{f}^{(t)} $ is separable, the solution of the subproblem~\eqref{prob:Q-update_inner} can be calculated in parallel for each independent variable $ q_i $.

The convergence of Algorithm~\ref{alg:SCA} is readily established in the SCA framework~\cite{yangUnifiedSuccessivePseudoconvex2017}. 
Specifically, due to the gradient consistency at the current point \(\vect{Q}^{(t)}\) between the original function $ f $ and the quadratic approximation $ \widetilde{f}^{(t)} $, when $ \widetilde{\vect{Q}}^{(t)} = \vect{Q}^{(t)} $, a stationary point of $ \widetilde{f}^{(t)} $ is achieved, which is also a stationary point and global optimum of the original function $ f $ in the set $\Sym_+^M \cap \set{T}^M$.
Otherwise, define \(\vect{\Delta}^{(t)} = \widetilde{\vect{Q}}^{(t)} - \vect{Q}^{(t)}\).
The difference \(\vect{\Delta}^{(t)}\) is a descent direction of $ f $ in the set $\Sym_+^M \cap \set{T}^M$. Thus, the matrix $\vect{Q}$ can be updated according to the rule
\begin{equation}
	\vect{Q}^{(t+1)} = \vect{Q}^{(t)} + \gamma^{(t)} \vect{\Delta}^{(t)}
\end{equation}
with a step size $\gamma^{(t)} \in (0, 1]$ chosen by Armijo backtracking line search, similar to that in Section~\ref{subsec:Q-update}, to ensure sufficient decrease of the original function $ f $. Since $\widetilde{\vect{Q}}^{(t)}$ lies in the original convex feasible set $ \Sym_+^M \cap \set{T}^M $, feasibility of $\vect{Q}^{(t+1)}$ is always maintained.


The proposed algorithm for problem~\eqref{prob:SI-SPARROW} in the SCA framework is outlined in Algorithm~\ref{alg:SCA}. 
The initialization strategy in Section~\ref{sec:ADMM_initialization} can also be applied to Algorithm~\ref{alg:SCA}. If the solution $\widehat{\vect{Q}}_2$ of the relaxed problem~\eqref{prob:ADMM_relax} is not PSD, i.e., not optimal for the original problem~\eqref{prob:SI-SPARROW}, then the following matrix in $\Sym_+^M \cap \set{T}^M$ near $\widehat{\vect{Q}}_2$ is chosen to be the initialization for $\vect{Q}$:
\begin{equation} \label{eq:SCA_initialization}
	\vect{Q}^{(0)} = \widehat{\vect{Q}}_2 - \lambda_{\min} (\widehat{\vect{Q}}_2) \vect{I}_M,
\end{equation}
where $\lambda_{\min}(\cdot)$ denotes the smallest eigenvalue.

\section{Generalization to Unobservable Sensors}
\label{sec:unobservable}

In this section, we briefly introduce a generalization of SI-SPARROW to the case where part of the sensors are unobservable.
We consider the same PCRA of $M$ sensors with the shift-invariance properties in~\eqref{eq:shift-invariance-prop} introduced in Section~\ref{sec:model}. Additionally, it is assumed that part of the sensors are unobservable, e.g., due to failure. In other scenarios, the unobservable sensors may be virtual sensors introduced for additional attractive geometric properties such as shift invariances or central symmetry required by Unitary ESPRIT, particularly for sparse arrays~\cite{pesaventoDirectionFindingPartly2002,yangRobustStatisticallyEfficient2023}.
Let $M' \leq M$ be the number of observable sensors with indices $\set{M}' = \{i_1, \ldots, i_{M'}\} $.
Define the selection matrix
\begin{equation}
	\vect{J}_{\set{M}'} = \begin{bmatrix}
		\vect{e}_{M,i_1} & \cdots & \vect{e}_{M,i_{M'}}
	\end{bmatrix} \in \Real^{M \times M'}.
\end{equation}
Then the matrix
	$\vect{Y}' = \vect{J}_{\set{M}'}^\Trans \vect{Y} \in \Compl^{M' \times N}$
contains the responses of the $M'$ observable sensors in $N$ time-slots.

To address the issue with unobservable sensors, a simple modification of SI-SPARROW introduced in Section~\ref{sec:problem} is to reconstruct or discard the part of the shift-invariance constraints that involve the unobservable sensors. This, however, leads to a significant degradation of the estimation quality.
Thus, we introduce a generalization of SI-SPARROW that, while using only the measurements $\vect{Y}'$ from the observable sensors for the data fidelity, preserves all the shift-invariance properties in~\eqref{eq:shift-invariance-prop} in the complete array.

When only the measurements in the matrix $\vect{Y}'$ are taken into consideration for the data fidelity, the DOA estimation problem based on the sparse model in~\eqref{eq:sparseModel} is then formulated as the following convex mixed-norm minimization problem:
\begin{equation}\label{prob:mixed-norm_unobservable}
	\min_{\vect{X} \in \Compl^{K \times N}} \quad \tfrac{1}{2} \norm{\vect{Y}' - \vect{J}_{\set{M}'}^\Trans \vect{A} \vect{X}}_\Frob^{2} + \lambda \sqrt{N}\norm{\vect{X}}_{2,1}.
\end{equation}
Similar to~\eqref{prob:mixed-norm}, problem~\eqref{prob:mixed-norm_unobservable} is equivalently reformulated as
\begin{equation}\label{prob:gridSPARROW_unobservable}
	\min_{{\vect{S}}\in \DiagSet_{+}^K} \quad \tr { \big(( \vect{J}_{\set{M}'}^\Trans \vect{A}\vect{S}\vect{A}^\Herm \vect{J}_{\set{M}'} + \lambda \vect{I}_{M'})^{\rm -1} \widehat{\vect{R}}' \big)}+\tr(\vect{S})
\end{equation}
with $\widehat{\vect{R}}' = \vect{Y}' (\vect{Y}')^\Herm / N$. Then we construct the following gridless relaxation of problem~\eqref{prob:gridSPARROW_unobservable} by replacing $\vect{ASA}^\Herm$ by a matrix $\vect{Q}$ that preserves the shift-invariance properties in~\eqref{eq:shift-invariance-constraint}:
\begin{equation} \label{prob:SI-SPARROW_unobservable}
	\min_{\vect{Q} \in \Sym_+^M \cap \set{T}^M} \ M \tr \big(( \vect{J}_{\set{M}'}^\Trans \vect{Q} \vect{J}_{\set{M}'} + \lambda \vect{I}_{M'})^{-1}\widehat{\vect{R}}' \big)+\tr (\vect{Q}).
\end{equation}
Likewise, problem~\eqref{prob:SI-SPARROW_unobservable} can be equivalently written as an SDP problem by the reformulation technique in Section~\ref{sec:problem} and solved by an interior-point solver. 
Alternatively, the two algorithms proposed in Sections~\ref{sec:ADMM} and~\ref{sec:SCA}, respectively, for the original SI-SPARROW problem~\eqref{prob:SI-SPARROW} can be easily adapted for solving problem~\eqref{prob:SI-SPARROW_unobservable} with unobservable sensors.
Due to space limitations, the details of the solution approaches for~\eqref{prob:SI-SPARROW_unobservable} are omitted and the application of this generalization of SI-SPARROW can be found in our subsequent work~\cite{liuBlindPhaseoffsetEstimation2024}.

\section{Simulation Results}
\label{sec:results}

In this section, we conduct numerical experiments on synthetic data to evaluate and analyze the performance of the developed SI-SPARROW method for a PCRA, as well as the complexity of the SDP approach and the two customized algorithms in Sections~\ref{sec:ADMM} and~\ref{sec:SCA}.
The SDP reformulation~\eqref{prob:SI-SPARROW-SDP-N} is modeled by CVX~\cite{grantCVXMatlabSoftware2020,grantGraphImplementationsNonsmooth2008} and solved by the interior-point solver MOSEK~\cite{mosekapsMOSEKOptimizationToolbox2023}.
As described in Section~\ref{subsec:SI-SPARROW}, from the reconstructed matrix $\widehat{\vect{Q}}$, the spatial frequencies are recovered by the MI-MD-ESPRIT method that is implemented using the simultaneous diagonalization~\cite{fuSimultaneousDiagonalizationSimilarity2006}, given the shift-invariance equations in~\eqref{eq:shift-invariance_sensor_x} and~\eqref{eq:shift-invariance_sensor_y} with the known intrasubarray displacements. 
The estimation error of the proposed method is compared to that of the conventional approach where the MI-MD-ESPRIT method is performed on the sample covariance matrix.
Unitary ESPRIT proposed in~\cite{haardtUnitaryESPRITHow1995}, and its multidimensional variant in~\cite{haardtSimultaneousSchurDecomposition1998}, exhibit an improved robustness to the source correlation compared to standard ESPRIT due to the incorporation of forward-backward averaging.
In principle, Unitary-ESPRIT cannot be applied to partly calibrated arrays as it relies on the centro-symmetric array configurations that cannot be ensured with the existence of intersubarray uncertainties.
Nevertheless, we choose centro-symmetric arrays in the simulations so as to include the multidimensional Unitary ESPRIT (MD-Unitary-ESPRIT) as a benchmark method for the performance comparison.
The MD-Unitary-ESPRIT method is also implemented using the simultaneous diagonalization but employs only a single shift-invariance equation per dimension.

Methods for partly calibrated arrays can readily be applied to the fully calibrated case.
For comparison, we also evaluate SI-SPARROW with full calibration where, from the solution of SI-SPARROW, the frequencies are recovered via MUSIC~\cite{schmidtMultipleEmitterLocation1986} using the complete array manifold. Similarly, the performance of SI-SPARROW in the fully calibrated case is compared to that of the conventional approach where MUSIC is performed on the sample covariance matrix.
\iflong
Additionally, the details of the derivation of the stochastic Cram\'er-Rao bound (CRB) in the partly calibrated case for the estimation of the spatial frequencies according to the signal model in~\eqref{eq:model} are given in Appendix~\ref{appendix:CRB}.
\fi

The results are averaged over $N_\text{R} \!=\! 1000$ Monte-Carlo trials.
The estimated frequencies $\widehat{\vect{\mu}}(n) \!=\! [(\widehat{\vect{\mu}}^x(n))^\Trans, (\widehat{\vect{\mu}}^y(n))^\Trans]^\Trans$ for $n \!=\! 1,\ldots,N_\text{R}$ are evaluated by the root-mean-square error (RMSE) with respect to the ground-truth $\vect{\mu}$ defined as
\begin{equation*}
	\text{RMSE} (\widehat{\bmu}) = \sqrt{\tfrac{1}{N_\text{S} N_\text{R}} \sum_{n=1}^{N_\text{R}} \sum_{i=1}^{N_\text{S}} \abs*{\hat{\mu}_i^x (n) - \mu_i^x}_\text{wa}^2 + \abs*{\hat{\mu}_i^y (n) - \mu_i^y}_\text{wa}^2},
\end{equation*}
where $ \hat{\mu}_i^x (n) $ and $ \hat{\mu}_i^y (n) $ are the estimates of the frequencies of source $i$ in the $x$-axis and $y$-axis, respectively, in the $ n $th trial, and $ \abs{\mu_1 - \mu_2}_\text{wa} = \min_{k \in \mathbb{Z}} \abs{\mu_1 - \mu_2 + 2 k \pi} $ is the wrap-around distance between two frequencies $\mu_1$ and $\mu_2$.
Experiments were conducted on a Linux PC with an Intel Core i7-7700 CPU and 32 GB RAM running MATLAB R2024a.

Simulations use $N_\text{S} = 2$ sources with spatial frequencies $\vect{\mu}^x = [0.5, 0.8]^\Trans$ and $\vect{\mu}^y = [1.5, 1.2]^\Trans$. The source waveforms follow a zero-mean complex normal distribution with unit variance and the correlation coefficient denoted by $\varphi$.
The PCRA consists of $4$ subarrays, each a uniform rectangular array with half-wavelength sensor spacing. If not further specified, each subarray contains $L_x = 4$ and $L_y = 4$ sensors in the two dimensions, respectively. 
The four subarrays are also placed uniformly in the two dimensions, specifically, at $(0,0)$, $(10,0)$, $(0,10)$, and $(10,10)$, and, without loss of generality, the phase-offsets $ \tau_p $ are set to be zero for all subarrays so that the complete array is centro-symmetric.
In the partly calibrated scenario only the sensor spacing within each subarray is assumed known. The SNR is calculated as $ 1/\sigma_n^2 $.

Although the normalized steering vector model is used in~\cite{steffensGridlessCompressedSensing2017}, the tuning rule for the regularization parameter proposed in~\cite{steffensGridlessCompressedSensing2017} can be readily adapted to the unnormalized steering vector model considered in this paper, which is $\lambda = \sqrt{M} \sigma_n (\sqrt{M/N}+1)$.
However, we choose a smaller value
	$\lambda = \sigma_n (\sqrt{M/N} + 1)$
to better balance the bias and variance to achieve a better overall RMSE performance in the inspected regions of SNR and sample size.

\subsection{Uncorrelated Source Signals}

\begin{figure}
	\centering
	\ref{legend2}
	
	\subfigure[RMSE vs. SNR for $N = 5$ snapshots \label{fig:varySNR_uncorr}]{\begin{tikzpicture}[inner ysep=1.5pt]
	\def\filepath{results/PCRA_general_intersubDis=10/4x4subarray/varySNR_N=5_uncorr_rmse.csv}
	\def\varName{SNR}
	
	\begin{axis}[%
		at={(0,0)},
		scale only axis,
		enlarge x limits={false},
    xmax=20,
		xlabel={SNR (dB)},
		ylabel={RMSE},
		ymode = log,
		axis background/.style={fill=white},
		xmajorgrids,
		ymajorgrids,
		legend style={legend cell align=left, align=left, draw=none, 
			font=\scriptsize, row sep=-2pt,
			legend pos=outer north east,
		},
		cycle multi list = {
			MATLABcolormarklist
		},
		legend to name=legend2,
		legend columns=3,
		]
		
		\addplot table [x index=0, y=SI-SPARROW(MOSEK)+MI-MD-ESPRIT] {\filepath};
		\addlegendentry{SI-SPARROW + MI-MD-ESPRIT};
		
		\addplot table [x=\varName, y=MI-MD-ESPRIT] {\filepath};
		\addlegendentry{MI-MD-ESPRIT};
		
		\addplot table [x=\varName, y=MD-Unitary-ESPRIT] {\filepath};
		\addlegendentry{MD-Unitary-ESPRIT};
		
		\addplot table [x index=0, y=SI-SPARROW+MUSIC] {\filepath};
		\addlegendentry{SI-SPARROW + MUSIC};
		
		\addplot table [x=\varName, y=MUSIC] {\filepath};
		\addlegendentry{MUSIC};
		
		\addplot[color=black, dashed] table [x=\varName, y=PCA-CRB] {\filepath};
		\addlegendentry{CRB (partly calibrated)};
		
		\addplot[color=black] table [x=\varName, y=CRB] {\filepath};
		\addlegendentry{CRB (fully calibrated)};
		

	\end{axis}

\end{tikzpicture}}
  \hfill
	\subfigure[RMSE vs. the number of snapshots for SNR $=0$ dB \label{fig:varyN_uncorr}]{\begin{tikzpicture}[inner ysep=1.5pt]
	\def\filepath{results/PCRA_general_intersubDis=10/4x4subarray/varyN_SNR=0_uncorr_rmse.csv}
	\def\varName{N}
	
	\begin{axis}[%
		at={(0,0)},
		scale only axis,
		enlarge x limits=false,
    xmax=1000,
		xlabel={Number of snapshots $N$},
		xmode = log,
		ylabel={RMSE},
		ymode = log,
		axis background/.style={fill=white},
		xmajorgrids,
		ymajorgrids,
		legend style={legend cell align=left, align=left, draw=none, 
			font=\tiny, row sep=-2pt,
			legend pos=outer north east,
		},
		cycle multi list = {
			MATLABcolormarklist
		},
		]
		
		\addplot table [x index=0, y=SI-SPARROW(MOSEK)+MI-MD-ESPRIT] {\filepath};
		\addlegendentry{SI-SPARROW + MI-MD-ESPRIT};
		
		\addplot table [x=\varName, y=MI-MD-ESPRIT] {\filepath};
		\addlegendentry{MI-MD-ESPRIT};
		
		\addplot table [x=\varName, y=MD-Unitary-ESPRIT] {\filepath};
		\addlegendentry{MD-Unitary-ESPRIT};
		
		\addplot table [x index=0, y=SI-SPARROW+MUSIC] {\filepath};
		\addlegendentry{SI-SPARROW + MUSIC};
		
		\addplot table [x=\varName, y=MUSIC] {\filepath};
		\addlegendentry{MUSIC};
		
		\addplot[color=black] table [x=\varName, y=CRB] {\filepath};
		\addlegendentry{CRB (fully calibrated)};
		
		\addplot[color=black, dashed] table [x=\varName, y=PCA-CRB] {\filepath};
		\addlegendentry{CRB}
		
		\legend{};
		
	\end{axis}
	
%
%
%
%
%
%
\end{tikzpicture}}
	\caption{Error performance for $N_\text{S} = 2$ uncorrelated sources with spatial frequencies $\vect{\mu}^x = [0.5, 0.8]^\Trans$ and $\vect{\mu}^y = [1.5, 1.2]^\Trans$. }
	\label{fig:rmse_uncorr}
\end{figure}

%
%

We first compare the error performance in the uncorrelated case for various choices of SNR and the number of snapshots, as shown in Fig.~\ref{fig:rmse_uncorr}.

In the partly calibrated case, when performed on the sample covariance matrix, MI-MD-ESPRIT shows a superior threshold performance in terms of both SNR and the number of snapshots compared to MD-Unitary-ESPRIT, due to using more shift invariances. When applied to the solution of SI-SPARROW, the RMSE of MI-MD-ESPRIT is further reduced in the region of low SNR or small number of snapshots, which also leads to an improved SNR threshold performance. 
All partly calibrated methods perform similarly in the asymptotic region.
Likewise, in the fully calibrated case, compared to MUSIC performed on the sample covariance matrix, MUSIC performed on the solution of SI-SPARROW shows similar asymptotic performance but has a superior threshold performance for both SNR and the number of snapshots.

\subsection{Correlated Source Signals}

\begin{figure}
	\centering
	\ref{legend2}
	
	\subfigure[RMSE vs. SNR for $N=5$ snapshots]{\begin{tikzpicture}[inner ysep=1.5pt]
	\def\filepath{results/PCRA_general_intersubDis=10/4x4subarray/varySNR_N=5_corr_rmse.csv}
	\def\varName{SNR}
	
	\begin{axis}[%
		at={(0,0)},
		scale only axis,
		enlarge x limits=false,
		xlabel={SNR (dB)},
		ylabel={RMSE},
		ymode = log,
		axis background/.style={fill=white},
		xmajorgrids,
		ymajorgrids,
		legend style={legend cell align=left, align=left, draw=white, 
			font=\tiny, row sep=-2pt,
			at={(0,0)}, anchor = south west},
		cycle multi list = {
			MATLABcolormarklist
		},
		]
		
		\addplot table [x index=0, y=SI-SPARROW(MOSEK)+MI-MD-ESPRIT] {\filepath};
		\addlegendentry{SI-SPARROW + MI-MD-ESPRIT};
		
		\addplot table [x=\varName, y=MI-MD-ESPRIT] {\filepath};
		\addlegendentry{MI-MD-ESPRIT};
		
		\addplot table [x=\varName, y=MD-Unitary-ESPRIT] {\filepath};
		\addlegendentry{MD-Unitary-ESPRIT};
		
		\addplot table [x index=0, y=SI-SPARROW+MUSIC] {\filepath};
		\addlegendentry{SI-SPARROW + MUSIC};
		
		\addplot table [x=\varName, y=MUSIC] {\filepath};
		\addlegendentry{MUSIC};
		
		\addplot[color=black] table [x=\varName, y=CRB] {\filepath};
		\addlegendentry{CRB};
		
		\addplot[color=black, dashed] table [x=\varName, y=PCA-CRB] {\filepath};
		\addlegendentry{PCA-CRB}
		
		\legend{};
		
	\end{axis}
\end{tikzpicture}} 
  \hfill
	\subfigure[RMSE vs. the number of snapshots for SNR $=0$ dB \label{fig:rmse_corr_varyN}]{\begin{tikzpicture}[inner ysep=1.5pt]
	\def\filepath{results/PCRA_general_intersubDis=10/4x4subarray/varyN_SNR=0_corr_rmse.csv}
	
	\begin{axis}[%
		at={(0,0)},
		scale only axis,
		enlarge x limits={false},
		xlabel={Number of snapshots $N$},
		xmode = log,
		ylabel={RMSE},
		ymode = log,
		axis background/.style={fill=white},
		xmajorgrids,
		ymajorgrids,
		legend style={legend cell align=left, align=left, draw=white, 
			row sep=-2pt,
			at={(0,0)}, anchor = south west},
		cycle multi list = {
			MATLABcolormarklist
		},
		]
		
		\addplot table [x=N, y=SI-SPARROW(MOSEK)+MI-MD-ESPRIT] {\filepath};
		\addlegendentry{SI-SPARROW + MI-MD-ESPRIT};
		
		\addplot table [x=N, y=MI-MD-ESPRIT] {\filepath};
		\addlegendentry{MI-MD-ESPRIT};
		
		\addplot table [x=N, y=MD-Unitary-ESPRIT] {\filepath};
		\addlegendentry{MD-Unitary-ESPRIT};
		
		\addplot table [x=N, y=SI-SPARROW+MUSIC] {\filepath};
		\addlegendentry{SI-SPARROW + MUSIC};
		
		\addplot table [x index = 0, y=MUSIC] {\filepath};
		\addlegendentry{MUSIC};
		
		\addplot[color=black, dashed] table [x=N, y=PCA-CRB] {\filepath};
		\addlegendentry{CRB (partly calibrated)};
		
		\addplot[color=black] table [x=N, y=CRB] {\filepath};
		\addlegendentry{CRB (fully calibrated)};
		
		\legend{};
		
	\end{axis}
\end{tikzpicture}}

	\caption{Error performance for $N_\text{S} = 2$ sources with correlation coefficient $\abs{\varphi} = 0.99$, and spatial frequencies $\vect{\mu}^x = [0.5, 0.8]^\Trans$ and $\vect{\mu}^y = [1.5, 1.2]^\Trans$. }
	\label{fig:rmse_corr}
\end{figure}

%

We then evaluate the error performance for correlated sources, with results shown in Fig.~\ref{fig:rmse_corr}.
The correlation coefficient between the two sources is chosen to be $\varphi = 0.99$.

We first compare the performance of the partly calibrated methods.
Surprisingly, Unitary ESPRIT does not provide a significantly improved error performance compared to standard ESPRIT, partly because MI-MD-ESPRIT utilizes more shift invariances than MD-Unitary-ESPRIT. 
The RMSE of both MI-MD-ESPRIT and MD-Unitary-ESPRIT is largely degraded compared to that in Fig.~\ref{fig:rmse_uncorr} with uncorrelated sources and does not approach the partly calibrated CRB. 
SI-SPARROW outperforms MI-MD-ESPRIT and MD-Unitary-ESPRIT in both asymptotic and non-asymptotic regions, especially in difficult scenarios, e.g., in the case with low SNR and/or with a limited number of snapshots. 
The RMSE of SI-SPARROW asymptotically converges to the partly calibrated CRB with the increase of SNR.
Although the same value of $ \lambda $ is used, the asymptotic bias of SI-SPARROW introduced by the regularization is more significant than that in Fig.~\ref{fig:rmse_uncorr} due to the high source correlation, as shown in the region of a large number of snapshots.

Similar to the partly calibrated case, in the fully calibrated case, the MUSIC method performed on the solution of SI-SPARROW exhibits a significantly improved error performance and a considerably lower threshold, compared to MUSIC performed on the sample covariance matrix. However, it also possesses an asymptotic bias with the increase of the number of snapshots due to the utilization of $\ell_{2,1}$-regularization.

\subsection{Computational Complexity}

In this subsection, we analyze the computational costs of the different methods for the SI-SPARROW problem~\eqref{prob:SI-SPARROW} and demonstrate them by numerical experiments.
Besides the two customized algorithms proposed in Sections~\ref{sec:ADMM} and~\ref{sec:SCA} and the benchmark SDP implementation via MOSEK, we additionally consider a single-loop ADMM algorithm obtained by adapting the ADMM-based approach in~\cite{yangGridlessSparseMethods2015,yangRobustStatisticallyEfficient2023} to an alternative SDP reformulation of problem~\eqref{prob:SI-SPARROW}, giving four solution approaches in total.
Applying Lemma 5 in~\cite{yangGridlessSparseMethods2015} to the expression~\eqref{eq:SI-SPARROW_obj_Y} of the objective function $f$, and following the same reformulation technique as in~\eqref{prob:SI-SPARROW-SDP-N}, problem~\eqref{prob:SI-SPARROW} can be recast as the SDP
\begin{subequations} \label{prob:SI-SPARROW2-SDP}
	\begin{align}
		\min_{\substack{\vect{G} \in \Compl^{M \times N'}, \\ \vect{q} \in \Compl^{I}, \, \vect{T} \in \Sym^{N'}}} & \quad \tr(T(\vect{q})) + \tfrac{M}{N'} \tr (\vect{T}) + \tfrac{M}{\lambda N'} \| \widehat{\vect{Y}} - \vect{G} \|_{\Frob}^{2} \\
		\st & \quad \begin{bmatrix} \vect{T} & \vect{G}^{\Herm} \\ \vect{G} & T(\vect{q}) \end{bmatrix} \succeq 0,
		 \quad q_1 \geq 0,
	\end{align}
\end{subequations}
with the auxiliary variables $\vect{G} \in \Compl^{M \times N'}$ and $\vect{T} \in \Sym^{N'}$.
Unlike the SDP reformulation~\eqref{prob:SI-SPARROW-SDP-N} solved by MOSEK, the semidefinite constraint in~\eqref{prob:SI-SPARROW2-SDP} already enforces the positive semidefiniteness of $T(\vect{q})$, so that ADMM applied directly to~\eqref{prob:SI-SPARROW2-SDP} admits a closed-form update for every primal variable and requires only a single outer loop, unlike the ADMM Algorithm~\ref{alg:ADMM} and the SCA Algorithm~\ref{alg:SCA}.
We refer to this algorithm as SDP-ADMM and include it in the comparison below.
However, the reduction of loop layers does not necessarily improve the scalability, since the considerably large number of auxiliary variables introduced by the SDP reformulation~\eqref{prob:SI-SPARROW2-SDP} affects both the per-iteration complexity and the number of iterations.

We first provide an analysis of the per-iteration complexity.
The ADMM Algorithm~\ref{alg:ADMM} and the SCA Algorithm~\ref{alg:SCA} both have a computational complexity of $\mathcal{O}(M^3)$ per inner iteration and an additional complexity of $\mathcal{O}(M^3)$ per outer iteration, which are independent of the sample size $N$, though the two algorithms arrive at this complexity differently.
In the ADMM Algorithm~\ref{alg:ADMM}, the $\vect{Q}$-update~\eqref{prob:Q-update} is solved by the SCA Algorithm~\ref{alg:innerSCA}, which constructs a quadratic approximation in each inner iteration.
Each inner iteration therefore requires inverting the $M \times M$ matrix $\vect{Q} + \lambda \vect{I}_M$ at a computational complexity of $\mathcal{O}(M^3)$, after which the resulting quadratic approximation is solved in closed form.
The subsequent Z-update~\eqref{prob:Z-update} in the outer iteration then applies the EVD-based projection $\opr{P}_{\Sym_+^M}$, likewise at $\mathcal{O}(M^3)$.
In the SCA Algorithm~\ref{alg:SCA}, the same $\mathcal{O}(M^3)$ matrix inversion is instead needed only once per outer iteration, to construct the quadratic approximation~\eqref{eq:f_approx}.
The resulting projection subproblem~\eqref{prob:quadraticApprox} is then solved by the inner ADMM iterations~\eqref{eq:ADMMiter_inner}, each of which again has a computational complexity of $\mathcal{O}(M^3)$ through the EVD-based projection.
Although SDP-ADMM performs a single loop, the SDP reformulation~\eqref{prob:SI-SPARROW2-SDP} introduces a considerably large number of auxiliary variables that are involved in the semidefinite constraint, which leads to a per-iteration complexity of $\mathcal{O}((M+N')^3)$ for the EVD-based projection $\opr{P}_{\Sym_+^{M+N'}}$.
The generic SDP reformulation~\eqref{prob:SI-SPARROW-SDP-N} solved via MOSEK also has a per-iteration complexity that grows with the dimension $M+N'$ of the semidefinite constraint, as well as with the number of optimization variables. See~\cite[Sec. 11.8.3]{boydConvexOptimization2004} for the detailed flop counts of an interior-point iteration.

%

\begin{figure}
	\centering
	\ref{legend3}
	
	\subfigure[Computation time vs. the number of snapshots \label{subfig:runtime_varyN}]{\begin{tikzpicture}[inner ysep=1.5pt]
	\def\filepath#1{results/PCRA_general_intersubDis=10/4x4subarray/varyN_SNR=0_corr_runtime#1.csv}
	\def\varName{N}
	
	\begin{axis}[%
		at={(0,0)},
		scale only axis,
		enlarge x limits=false,
    xmax=1000,
		xlabel={Number of snapshots $N$},
		xmode = log,
    extra x ticks={64},
    extra x tick labels={$ M $},
		ylabel={Computation time (s)},
		ymode = log,
		axis background/.style={fill=white},
		xmajorgrids,
		ymajorgrids,
		legend style={
		},
		cycle multi list = {
			MATLABcolormarklist
		},
		legend to name=legend3,
		legend columns=3,
		]
		
		\addplot table [x=\varName, y=SI-SPARROW(MOSEK)+MI-MD-ESPRIT] {\filepath1};
		\addlegendentry{SI-SPARROW (MOSEK)};
		
		\pgfplotsset{cycle list shift=2};
		
		\addplot table [x=\varName, y=SI-SPARROW(ADMM)+MI-MD-ESPRIT] {\filepath1};
		\addlegendentry{SI-SPARROW (ADMM)};
		
		\pgfplotsset{cycle list shift=3};
		
		\addplot table [x=\varName, y=SI-SPARROW(SCA)+MI-MD-ESPRIT] {\filepath1};
		\addlegendentry{SI-SPARROW (SCA)};
		
		\addplot table [x=\varName, y=SI-SPARROW(SDP-ADMM)+MI-MD-ESPRIT] {\filepath1};
		\addlegendentry{SI-SPARROW (SDP-ADMM)};

    \addplot[color=black, dashed] table [x=\varName, y=SI-SPARROW(ADMM)+MI-MD-ESPRIT] {\filepath3};
    \addlegendentry{SCA Algorithm~\ref{alg:innerSCA} for solving relaxed problem~\eqref{prob:ADMM_relax}};

		\pgfplotsset{cycle list shift=-4};

		\addplot table [x=\varName, y=MI-MD-ESPRIT] {\filepath2};
		\addlegendentry{MI-MD-ESPRIT};

		\addplot table [x=\varName, y=MD-Unitary-ESPRIT] {\filepath2};
		\addlegendentry{MD-Unitary-ESPRIT};

		\pgfplotsset{cycle list shift=-3};

		\addplot table [x=\varName, y=MUSIC] {\filepath2};
		\addlegendentry{MUSIC};	
		
		
	\end{axis}
\end{tikzpicture}}
	\hfill
	\subfigure[Number of outer iterations vs. the number of snapshots \label{}]{\begin{tikzpicture}[inner ysep=1.5pt]
	\def\filepath#1{results/PCRA_general_intersubDis=10/4x4subarray/varyN_SNR=0_corr_#1.csv}
	\def\varName{N}
	
	\begin{axis}[%
		at={(0,0)},
		scale only axis,
		enlarge x limits={false},
    xmax=1000,
		xlabel={Number of snapshots $N$},
		xmode = log,
    extra x ticks={64},
    extra x tick labels={$ M $},
    ylabel={Number of outer iterations},
		ymode = log,
		axis background/.style={fill=white},
		xmajorgrids,
		ymajorgrids,
		legend style={legend cell align=left, align=left, draw=white, 
			row sep=-2pt,
			at={(0,0)}, anchor = south west},
		cycle multi list = {
			MATLABcolormarklist
		},
		]
		
		
		\pgfplotsset{cycle list shift=3};
		
    \addplot table [x=\varName, y=SI-SPARROW(ADMM)+MI-MD-ESPRIT] {\filepath{nIter}};
		\addlegendentry{SI-SPARROW (ADMM)};
		
		\pgfplotsset{cycle list shift=4};

    \addplot table [x=\varName, y=SI-SPARROW(SCA)+MI-MD-ESPRIT] {\filepath{nIter}};
		\addlegendentry{SI-SPARROW (SCA)};

    \addplot table [x=\varName, y=SI-SPARROW(SDP-ADMM)+MI-MD-ESPRIT] {\filepath{nIter}};
		\addlegendentry{SI-SPARROW (SDP-ADMM)};

		%

		\legend{};
		
	\end{axis}
\end{tikzpicture}}
  \hfill
	\subfigure[Total number of inner iterations vs. the number of snapshots \label{}]{\begin{tikzpicture}[inner ysep=1.5pt]
	\def\filepath#1{results/PCRA_general_intersubDis=10/4x4subarray/varyN_SNR=0_corr_#1.csv}
	\def\varName{N}
	
	\begin{axis}[%
		at={(0,0)},
		scale only axis,
		enlarge x limits={false},
    xmax=1000,
		xlabel={Number of snapshots $N$},
		xmode = log,
    extra x ticks={64},
    extra x tick labels={$ M $},
    ylabel={Total number of inner iterations},
		ymode = log,
		axis background/.style={fill=white},
		xmajorgrids,
		ymajorgrids,
		legend style={legend cell align=left, align=left, draw=white, 
			row sep=-2pt,
			at={(0,0)}, anchor = south west},
		cycle multi list = {
			MATLABcolormarklist
		},
		]
		
		
		\pgfplotsset{cycle list shift=3};
		
    \addplot table [x=\varName, y=SI-SPARROW(ADMM)+MI-MD-ESPRIT] {\filepath{totalInnerIter}};
		\addlegendentry{SI-SPARROW (ADMM)};
		
		\pgfplotsset{cycle list shift=4};
		
    \addplot table [x=\varName, y=SI-SPARROW(SCA)+MI-MD-ESPRIT] {\filepath{totalInnerIter}};
		\addlegendentry{SI-SPARROW (SCA)};
		
    \addplot table [x=\varName, y=SI-SPARROW(SDP-ADMM)+MI-MD-ESPRIT] {\filepath{totalInnerIter}};
		\addlegendentry{SI-SPARROW (SDP-ADMM)};

		\pgfplotsset{cycle list shift=1};

    \addplot[color=black, dashed] table [x=\varName, y=SI-SPARROW(ADMM)+MI-MD-ESPRIT] {\filepath{initIter}};
    \addlegendentry{SCA Algorithm~\ref{alg:innerSCA} for solving relaxed problem~\eqref{prob:ADMM_relax}};
		
		\legend{};
		
		\legend{};
		
	\end{axis}
\end{tikzpicture}}
	\caption{Computational complexity for $N_\text{S} \!=\! 2$ sources with correlation coefficient $\abs{\varphi} \!=\! 0.99$, $\vect{\mu}^x = [0.5, 0.8]^\Trans$ and $\vect{\mu}^y = [1.5, 1.2]^\Trans$, and SNR $=0$ dB. }
	\label{fig:runtime}
\end{figure}

The overall computational complexity is equal to the product of the per-iteration complexity and the number of iterations, which is reflected in the empirical results in Fig.~\ref{fig:runtime}.
We keep the scenario from the previous section with correlation coefficient $\varphi = 0.99$.
The convexity of problem~\eqref{prob:SI-SPARROW} ensures that all the four solution approaches converge to the global optimum.
However, as they are established on different reformulations of problem~\eqref{prob:SI-SPARROW}, it is difficult to compare their convergence rates by tracking the objective function values.
Instead, we properly choose the termination tolerances for the three customized algorithms so that they achieve similar precisions on the RMSE as that of the benchmark MOSEK in Fig.~\ref{fig:rmse_corr_varyN} and then compare their computational costs.

The same initialization strategy described in Section~\ref{sec:ADMM_initialization} is applied to all three customized algorithms for SI-SPARROW. That is, we first solve the relaxed problem~\eqref{prob:ADMM_relax} without the PSD constraint by using the SCA Algorithm~\ref{alg:innerSCA}. To better compare the computational complexities of the three customized algorithms, we show the computational costs of the common initialization step separately from the main algorithms.
Apart from the computation time, in Fig.~\ref{fig:runtime}, we also report the number of iterations for the three customized algorithms, as well as the sum of all inner iterations in the double-loop ADMM Algorithm~\ref{alg:ADMM} and SCA Algorithm~\ref{alg:SCA}.

The computational cost of MUSIC is dominated by the grid search over the 2D MUSIC spectrum, which depends mainly on the grid size.
Both MI-MD-ESPRIT and MD-Unitary-ESPRIT admit closed-form solutions given the simultaneous diagonalization and, hence, they have the lowest computational cost, which is also independent of the number of snapshots.

Then we compare the four solution approaches for SI-SPARROW.
Their differences in computational complexity result mainly from the different problem formulations that they exploit.
First notably, the solution of the relaxed problem~\eqref{prob:ADMM_relax} becomes closer to PSD as the sample size increases.
As a result, the number of outer iterations in the ADMM Algorithm~\ref{alg:ADMM} and the SCA Algorithm~\ref{alg:SCA} significantly reduces.
On the contrary, since the SDP-ADMM algorithm is established on a SDP reformulation~\eqref{prob:SI-SPARROW2-SDP}, it cannot benefit from the above property of the solution of the relaxed problem~\eqref{prob:ADMM_relax}.
Moreover, in the undersampled case, the number of columns $N'$ of the compressed measurement matrix $\widehat{\vect{Y}}$ equals the sample size $N$, so that the number of auxiliary variables in~\eqref{prob:SI-SPARROW2-SDP} grows with $N$, which further leads to a significant increase in the number of iterations of SDP-ADMM.
The growth of $N'$ also enlarges the semidefinite constraint in~\eqref{prob:SI-SPARROW2-SDP}, so that the per-iteration complexity $\mathcal{O}((M+N')^3)$ of SDP-ADMM increases with $N$ as well, whereas the per-iteration complexity $\mathcal{O}(M^3)$ of the ADMM Algorithm~\ref{alg:ADMM} and the SCA Algorithm~\ref{alg:SCA} is independent of $N$.
Since both factors in this product grow with the sample size, the computation time of the SDP-ADMM algorithm increases dramatically with the increase of the sample size. A similar behavior is observed for the SDP implementation via MOSEK.
By avoiding an additional SDP reformulation, the ADMM Algorithm~\ref{alg:ADMM} and the SCA Algorithm~\ref{alg:SCA} obtain an improved scalability with respect to the sample size.

In the oversampled case, the solution of the relaxed problem~\eqref{prob:ADMM_relax} is often already PSD and, hence, optimal for the original SI-SPARROW problem~\eqref{prob:SI-SPARROW}. 
In consequence, the computational costs of the three customized algorithms for SI-SPARROW suddenly drop to zero as they are no longer performed.
Moreover, the dimensionality reduction in~\eqref{eq:measurementCompress} caps $N'$ at $M$, so that the per-iteration complexity of the SDP implementation via MOSEK, and hence its computation time, stops growing as the sample size further increases.

In summary, each customized algorithm demonstrates its advantages in different scenarios.

  \textit{SDP-ADMM}: It has the lowest computational cost in the case of a small number of snapshots.

  \textit{ADMM Algorithm~\ref{alg:ADMM}}:
    By avoiding an additional SDP reformulation, the ADMM Algorithm~\ref{alg:ADMM} has a better scalability with respect to the sample size than SDP-ADMM.

  \textit{SCA Algorithm~\ref{alg:SCA}}:
    Similar to the ADMM Algorithm~\ref{alg:ADMM}, the SCA Algorithm~\ref{alg:SCA} also scales better with the increase of the sample size than SDP-ADMM.
    Although the SCA Algorithm~\ref{alg:SCA} shows a higher computational cost than the ADMM Algorithm~\ref{alg:ADMM}, it has the advantage that the quadratic approximation converts the original problem to a simple projection onto the set of structured PSD matrices. 
  Special cases of this projection problem with, e.g., a Toeplitz structure, which exists in the case of ULAs, have been extensively studied~\cite{suffridgeApproximationHermitianPositive1993}.
  Any future advances in developing new techniques for this projection problem with a particular structure or the generalized shift-invariant structure can be immediately used in Algorithm~\ref{alg:SCA} to further reduce the computational cost.

\section{Conclusion}
\label{sec:conclusion}

We consider the 2D DOA estimation problem from multiple measurement vectors (MMVs) using a partly calibrated rectangular array (PCRA). 
Whereas the existing gridless methods are restricted to the case of uniform and fully calibrated array geometries, we present a gridless sparsity-based formulation that leverages the shift invariances, termed SI-SPARROW, for the DOA estimation in PCRAs.
Numerical simulations demonstrate the robustness of our proposed method to the source correlation, compared to the conventional partly calibrated methods, where the ESPRIT-like methods are performed on the sample covariance matrix.

In contrast to the SDP reformulation approach in~\cite{steffensCompactFormulationEll2018}, we develop two customized algorithms under the ADMM and successive convex approximation (SCA) frameworks for the established SI-SPARROW problem.
Each of them has merit in terms of the computational complexity in different scenarios.

In addition, since no calibration information is involved in the gridless SI-SPARROW formulation, this technique can also be used in the fully calibrated case as a low-cost decorrelation step to improve the robustness of subspace-based methods, e.g., MUSIC, to the source correlation. 

\appendix
\numberwithin{equation}{section}

\section{Second-Order Taylor Expansions of $ f $ in~\eqref{prob:SI-SPARROW}}
\label{appendix:Taylor}
In this section, we derive the gradient as well as the first- and second-order Taylor expansions of the function $f$ in~\eqref{prob:SI-SPARROW}.

In this paper, we only discuss the gradient of a real-valued function. 
The general definition of a gradient given in~\eqref{eq:partialGradient} is derived from the first-order Taylor expansion expressed with the Wirtinger derivatives. For a real-valued function $f$ with complex-valued arguments $\vect{x}$, we have $\left(\frac{\partial f}{\partial \vect{x}}\right)^* = \frac{\partial f}{\partial \vect{x}^*}$. Hence, the first-order Taylor expansion of $f$ around a point $\vect{x}^o$ is
\begin{align*}
  f(\vect{x}) &\approx f(\vect{x}^o) + \left(\tfrac{\partial f (\vect{x}^o)}{\partial \vect{x}}\right)^{\Trans} (\vect{x} - \vect{x}^o) + \left(\tfrac{\partial f(\vect{x}^o)}{\partial \vect{x}^*}\right)^{\Trans} (\vect{x} - \vect{x}^o)^* \\
	&= f(\vect{x}^o) + \Re \left( \left(2 \tfrac{\partial f (\vect{x}^o)}{\partial \vect{x}^*}\right)^\Herm (\vect{x} - \vect{x}^o) \right),
\end{align*}
which implies that the gradient of $f$ at $\vect{x}^o$ is \begin{equation} \label{eq:def_grad}
	\nabla_\vect{x} f (\vect{x}^o) = 2 \tfrac{\partial f (\vect{x}^o)}{\partial \vect{x}^*}.
\end{equation}

However, the expression in~\eqref{eq:def_grad} cannot be applied to $f$ in~\eqref{prob:SI-SPARROW} as $\vect{Q}$ admits a Hermitian structure.
To derive the gradient of $f$ in~\eqref{prob:SI-SPARROW} with respect to a Hermitian matrix $\vect{Q}$, we first establish the first-order Taylor expansion of $f$. For an unstructured matrix $\vect{Q} \in \Compl^{M \times M}$, $f$ is no longer real-valued and its Wirtinger derivatives are given by~\cite{hjorungnesComplexValuedMatrixDifferentiation2007}
\begin{align}
	\tfrac{\partial f (\vect{Q})}{\partial \vect{Q}} &= -M \big[(\vect{Q} + \lambda \vect{I}_M)^{-1} \widehat{\vect{R}} (\vect{Q} + \lambda \vect{I}_M)^{-1}\big]^\Trans + \vect{I}_M, \nonumber \\
	\tfrac{\partial f (\vect{Q})}{\partial \vect{Q}^*} &= \vect{0}.
\end{align}
With the above Wirtinger derivatives, the first-order Taylor expansion of $f$ around a given point $\vect{Q}^{(l)}$ can be written as
\begin{align}
  f(\vect{Q}) &\approx f(\vect{Q}^{(l)}) + \begin{matrix} \sum_{i=1}^M \sum_{j=1}^M \tfrac{\partial f (\vect{Q}^{(l)})}{\partial q_{i,j}} \big(q_{i,j}-q_{i,j}^{(l)} \big) \end{matrix} \nonumber \\
	&= f(\vect{Q}^{(l)}) + \tr \left( \left(\tfrac{\partial f (\vect{Q}^{(l)})}{\partial \vect{Q}}\right)^\Trans \vect{\Delta} \right), \label{eq:linearTaylor}
\end{align}
where $\vect{\Delta} = \vect{Q} - \vect{Q}^{(l)}$. When $\vect{Q}$ is limited to be Hermitian, both terms in~\eqref{eq:linearTaylor} become real-valued and the second term is the standard inner product of $\left(\frac{\partial f(\vect{Q}^{(l)})}{\partial \vect{Q}}\right)^*$ and $\vect{\Delta}$ in the complex space. Hence, it can be identified that the gradient of $f$ with respect to a Hermitian matrix $\vect{Q}$ at $\vect{Q}^{(l)}$ is
\begin{equation} \label{eq:gradf_Q_2}
	\nabla_{\vect{Q}} f (\vect{Q}^{(l)}) = \left(\tfrac{\partial f(\vect{Q}^{(l)})}{\partial \vect{Q}}\right)^*. 
\end{equation}

\begin{figure*}[pos=t]
	\centering
	\begin{align}
		\nabla_{\vect{Q}} f (\vect{Q}) &= - M (\vect{W+\Delta})^{-1} \cdot \widehat{\vect{R}} \cdot (\vect{W+\Delta})^{-1} + \vect{I}_M \nonumber \\
		&= - M \vect{W}^{-\frac{1}{2}} \left( \vect{I}_M + \vect{W}^{-\frac{1}{2}} \vect{\Delta W}^{-\frac{1}{2}} \right)^{-1} \vect{W}^{-\frac{1}{2}} \cdot \widehat{\vect{R}} \cdot \vect{W}^{-\frac{1}{2}} \left( \vect{I}_M + \vect{W}^{-\frac{1}{2}} \vect{\Delta W}^{-\frac{1}{2}} \right)^{-1} \vect{W}^{-\frac{1}{2}} + \vect{I}_M \nonumber \\
		&\approx - M \vect{W}^{-\frac{1}{2}} \left( \vect{I}_M - \vect{W}^{-\frac{1}{2}} \vect{\Delta W}^{-\frac{1}{2}} \right) \vect{W}^{-\frac{1}{2}} \cdot \widehat{\vect{R}} \cdot \vect{W}^{-\frac{1}{2}} \left( \vect{I}_M - \vect{W}^{-\frac{1}{2}} \vect{\Delta W}^{-\frac{1}{2}} \right) \vect{W}^{-\frac{1}{2}} + \vect{I}_M \nonumber \\
		&= \underbrace{- M \vect{W}^{-1}\widehat{\vect{R}} \vect{W}^{-1} + \vect{I}_M}_{\nabla_\vect{Q}f(\vect{Q}^{(l)})} + M (\vect{W}^{-1} \vect{\Delta W}^{-1} \widehat{\vect{R}} \vect{W}^{-1} + \vect{W}^{-1} \widehat{\vect{R}} \vect{W}^{-1} \vect{\Delta W}^{-1}) - \underbrace{ M \vect{W}^{-1} \vect{\Delta W}^{-1} \widehat{\vect{R}} \vect{W}^{-1} \vect{\Delta W}^{-1} }_{o(\norm{\vect{\Delta}}_\Frob)} \nonumber \\
		&\approx \nabla_{\vect{Q}} f (\vect{Q}^{(l)}) + M (\vect{W}^{-1} \vect{\Delta W}^{-1} \widehat{\vect{R}} \vect{W}^{-1} + \vect{W}^{-1} \widehat{\vect{R}} \vect{W}^{-1} \vect{\Delta W}^{-1}) \label{eq:gradf_Q_approx}
	\end{align}
  \hrule
\end{figure*}

Then, following the same procedure as in~\cite[Appendix A.4.3]{boydConvexOptimization2004}, we can obtain the second-order Taylor expansion of \(f\) in~\eqref{prob:SI-SPARROW} around \(\vect{Q}^{(l)}\) by first deriving the first-order Taylor expansion of its gradient. For a point \(\vect{Q}\) near \(\vect{Q}^{(l)}\), define \(\vect{\Delta} = \vect{Q} - \vect{Q}^{(l)}\) and \(\vect{W} = \vect{Q}^{(l)} + \lambda \vect{I}_M\). 
For the gradient $ \nabla_{\vect{Q}} f $ in~\eqref{eq:gradf_Q_2}, its first-order Taylor expansion around $\vect{Q}^{(l)}$ can be expressed in~\eqref{eq:gradf_Q_approx} by using the first-order approximation \((\vect{I+A})^{-1} \approx \vect{I-A}\) for \(\norm{\vect{A}}_\Frob\) being small.
Then the second-order Taylor expansion of \(f\) in~\eqref{eq:fTaylor} is obtained by integrating the first-order Taylor expansion of its gradient.

\section{Stochastic Cram\'er-Rao Bound for PCRAs}
\label{appendix:CRB}

This section details the calculation of the stochastic Cram\'er-Rao Bound (CRB) for the estimation of the spatial frequencies $\vect{\mu}$ in PCRAs under the signal model~\eqref{eq:model}, following the approach in~\cite{seeDirectionofarrivalEstimationPartly2004}. For simplicity, the steering matrix $\vect{A}(\vect{\mu}, \vect{\alpha})$ in~\eqref{eq:model} is denoted by $\vect{A}$ in this section.

The stochastic model, also referred to as the unconditional model assumption (UMA)~\cite{stoicaPerformanceStudyConditional1990,stoicaStochasticCRBArray2001}, assumes that the source waveforms $\vect{\Psi}$ are stochastic. The waveform vector $\vect{\psi}_n$ at each time instant i.i.d. follows the Gaussian distribution
	$\mathcal{CN} (\vect{0}, \vect{P})$,
where $\vect{P} \in \Compl^{N_\text{S} \times N_\text{S}}$ is the unknown source covariance matrix. Then, each snapshot $\vect{y}_n$ i.i.d. follows the Gaussian distribution
\begin{equation} \label{eq:stochasticModel}
	\vect{y}_n \sim \mathcal{CN} (\vect{0}, \vect{R}) \quad \text{with} \quad \vect{R} = \vect{APA}^\Herm + \sigma_n^2 \vect{I}_M.
\end{equation}

To simplify the calculation of the CRB, we ignore the relation in~\eqref{eq:uncalibratedSteerVec} 
and denote the vector $\vect{h}(\mu^x_i,\mu^y_i,\vect{\alpha})$ as $\vect{h}_i$.
Note that the constraint~\eqref{eq:shift-invariance-constraint_intersubarray} derived from the intersubarray shift invariances is established on the assumption that $ \vect{h}_i $ contains only unit-modulus entries. In contrast to~\cite{seeDirectionofarrivalEstimationPartly2004}, which directly considers the entries of the vectors $ \vect{h}_i $ as the parameters to be recovered, we employ the following reparameterization:
\begin{equation} \label{}
  \vect{h}_i = \begin{bmatrix}
    1, \euler^{\imagunit \xi_{2,i}}, \ldots, \euler^{\imagunit \xi_{P,i}}
  \end{bmatrix}^\Trans
\end{equation}
for $ i = 1,\ldots,N_\text{S} $, to preserve the information that $ \vect{h}_i $ contains unit-modulus entries.
Thus, we summarize all the relevant unknown parameters in a real-valued 
$(P+1) N_\text{S} \times 1$
vector
\begin{equation}
	\vect{\eta} = \big[
  {(\vect{\mu}^x)}^\Trans, {(\vect{\mu}^y)}^\Trans, \vect{\xi}_2^\Trans, \ldots, \vect{\xi}_{P}^\Trans \big]^\Trans,
\end{equation}
where
	$\vect{\xi}_p = [\xi_{p,1}, \ldots, \xi_{p,N_\text{S}}]^\Trans$ for $p = 2,\ldots,P$. 

Apart from $ \vect{\eta} $, the UMA model~\eqref{eq:stochasticModel} also involves the noise variance $\sigma_n^2$ and the source covariance matrix $\vect{P}$ as unknown parameters.
However, as shown in~\cite{stoicaStochasticCRBArray2001,stoicaPerformanceStudyConditional1990}, only the parameters in $\vect{\eta}$ are involved in the calculation of the block of the CRB matrix corresponding to the spatial frequencies $\vect{\mu}$, which we are interested in.
By concentrating the problem with respect to the nuisance parameters $\sigma_n^2$ and $\vect{P}$, the entries of the inverse of the $(P+1)N_\text{S} \times (P+1)N_\text{S}$ $\vect{\eta}$-block of the CRB matrix, denoted by $\text{CRB}(\vect{\eta})$, can be expressed as
\begin{equation} \label{eq:inverseCRB}
	\left[\text{CRB}^{-1} (\vect{\eta})\right]_{k,l} = \tfrac{2N}{\sigma_n^2} \Re \left( \tr \left(\vect{U} \tfrac{\partial \vect{A}^\Herm}{\partial \eta_k} \vect{\Pi}_{\vect{A}}^\perp \tfrac{\partial \vect{A}}{\partial \eta_l}\right)\right),
\end{equation}
where $\vect{\Pi}_{\vect{A}}^\perp = \vect{I} - \vect{A} (\vect{A}^\Herm \vect{A})^{-1} \vect{A}^\Herm$ is the orthogonal projector onto the orthogonal complement of the range of $\vect{A}$ and
\begin{equation}
	\vect{U} = \vect{PA}^\Herm \vect{R}^{-1} \vect{AP} = \vect{P} (\vect{A}^\Herm \vect{AP} + \sigma_n^2 \vect{I})^{-1} \vect{A}^\Herm \vect{AP}.
\end{equation}
Omitting the intermediate calculations from~\eqref{eq:inverseCRB}, we can write the $\vect{\eta}$-block of the CRB matrix in a compact form as
\begin{align}
  \text{CRB} (\vect{\eta}) =& \tfrac{\sigma_n^2}{2N} \left[\Re \left( (\vect{D}^\Herm \vect{\Pi}_{\vect{A}}^\perp \vect{D}) \odot (\vect{11}^\Trans \otimes \vect{U}^\Trans)\right)\right]^{-1} \\
  \text{with} \quad \vect{D} =& \big[ \vect{D}_{\vect{\mu}^x}, \vect{D}_{\vect{\mu}^y},\vect{D}_{\vect{\xi}_2},\ldots,\vect{D}_{\vect{\xi}_{P}} \big], \nonumber \\
	\vect{D}_{\vect{\mu}^x} =& \begin{bmatrix}
		\frac{\partial \vect{a}(\mu_1^x,\mu_1^y)}{\partial \mu_1^x}, \ldots, \frac{\partial \vect{a}(\mu_{N_\text{S}}^x,\mu_{N_\text{S}}^y)}{\partial \mu_{N_\text{S}}^x}
	\end{bmatrix}, \nonumber \\
	\vect{D}_{\vect{\mu}^y} =& \begin{bmatrix}
		\frac{\partial \vect{a}(\mu_1^x,\mu_1^y)}{\partial \mu_1^y}, \ldots, \frac{\partial \vect{a}(\mu_{N_\text{S}}^x,\mu_{N_\text{S}}^y)}{\partial \mu_{N_\text{S}}^y}
	\end{bmatrix}, \nonumber \\
  \vect{D}_{\vect{\xi}_p} =& \begin{bmatrix}
		\frac{\partial \vect{a} (\mu^x_1,\mu^y_1)}{\partial \xi_{p,1}},\ldots,\frac{\partial \vect{a} (\mu^x_{N_\text{S}}, \mu^y_{N_\text{S}})}{\partial \xi_{p,N_\text{S}}}
	\end{bmatrix} 
	 \quad \text{for } p=2,\ldots,P. \nonumber
\end{align}
The square root of the average of the diagonal entries corresponding to the frequencies $\vect{\mu}$ in the CRB matrix provides a lower bound for the RMSE in the estimation of $\vect{\mu}$ and it is used as a reference for the performance evaluation in Section~\ref{sec:results}.

\section*{Acknowledgements}
Funding: This work was supported in part by the DFG PRIDE Project PE 2080/2-1 and in part by the EXPRESS II Project within the DFG priority program CoSIP (DFG-SPP 1798). The funding sources had no involvement in the study design, in the collection, analysis and interpretation of data, in the writing of the article, or in the decision to submit it for publication.

\bibliographystyle{elsarticle-num-names}
\bibliography{refsDOA}

\end{document}